\documentclass[conference]{IEEEtran}
\IEEEoverridecommandlockouts
\usepackage{cite}
\usepackage{amsmath,amssymb,amsfonts}
\usepackage{algorithmic}
\usepackage{graphicx}
\usepackage{textcomp}
\usepackage{xcolor}
\usepackage{multirow}
\def\BibTeX{{\rm B\kern-.05em{\sc i\kern-.025em b}\kern-.08em
    T\kern-.1667em\lower.7ex\hbox{E}\kern-.125emX}}

\newcommand{\cut}[1]{{}}
\newcommand{\SpendableStore}{\textit{SpendableStore}}
\newcommand{\SSInstance}{\textit{SS-Instance}}
\newcommand{\SpendableDO}{\textit{SDO}}
\newcommand{\FNSI}{\textit{FNSI}}
\newcommand{\UTXO}{\textit{UTxO}}
\newcommand{\TXO}{\textit{TxO}}
\newcommand{\oVersion}{\textit{OV}}
\newcommand{\SSTx}{\textit{SS-Tx}}

\newcommand{\revise}{}
\newcommand{\removed}[1]{}

\usepackage{enumitem}
\usepackage[normalem]{ulem}
\usepackage{graphicx}
\usepackage{caption}
\usepackage{url}
\usepackage{hyperref}

\begin{document}
\captionsetup{skip=3pt}

\title{SpendableStore: A UTXO-based\\Decentralized Data Store
\thanks{This research is partly supported by the NSF under grant SaTC-2245372.}
}

\author{\IEEEauthorblockN{Yinan Zhou}
\IEEEauthorblockA{\textit{University of California, Irvine} \\
\textit{yinanz17@uci.edu}}
\and
\IEEEauthorblockN{Faisal Nawab}
\IEEEauthorblockA{\textit{University of California, Irvine} \\
\textit{nawabf@uci.edu}}
}

\maketitle

\begin{abstract}
The literature on blockchain-based databases is divided into permissioned blockchains and permissionless account-based blockchains. However, the former is not fully decentralized, and the latter suffers from challenges in scalability and practicality. We propose SpendableStore, a hybrid on/off-chain database that operates on top of permissionless UTXO-based blockchains as a novel approach to the problem of data decentralization. Our design integrates atomic data units into individual UTXOs to create a new blockchain concept called Spendable Data Objects that perform traditional CRUD operations. The integrity, immutability, and ownership of these Spendable Data Objects are safeguarded directly by the blockchain peer nodes, thus constraining the power of database administrators to achieve true data decentralization. We further support database transactions and propose an isolation mechanism called Future Now Snapshot Isolation to reason about transactional correctness in SpendableStore. We performed experiments on a major blockchain's Mainnet and observed up to 16x better throughput compared to a state-of-the-art blockchain-based database.
\end{abstract}


\section{Introduction}
\label{section:Introduction}
The creation and implementation of the Bitcoin~\cite{nakamoto2008bitcoin} blockchain protocol became a breakthrough in decentralized systems. Its use of the Proof-of-Work mechanism was a novel solution to the Byzantine Generals Problem~\cite{lamport2019byzantine}. It offers a transparent, immutable, decentralized ledger in a fully open network. Bitcoin demonstrates a new approach to achieving desirable system properties such as decentralization, tamper resistance, irrefutability, and anonymity~\cite{korth2021notes}.

Blockchain-related databases are one line of research with various approaches (details in Section~\ref{section: Related Works}). Many works mainly aim to utilize the verifiability property of the blockchain rather than its decentralization property. Therefore, they are built on less decentralized options like permissioned, federated blockchains, blockchains that are sharded~\cite{zamani2018rapidchain, Fang2022PeloPartition}, sidechains~\cite{singh2020sidechain}, or storing data in centralized off-chain nodes~\cite{xu2021slimchain}. The decentralization aspect of these systems is often traded off for performance gain.

Systems that are designed to operate on decentralized blockchain networks mainly use Ethereum Smart Contracts~\cite{tan2020latte, bragagnolo2018smartinspect}. Ethereum follows an account-based~\cite{zahnentferner2018chimeric, brunjes2020utxo} design where each contract account is a stateful software program that sequentially executes all update requests. On the other hand, Bitcoin has been used mainly for decentralized monetary transactions and value stores ~\cite{baur2021volatility, kubat2015virtual, mattke2020cryptocurrency}. However, we found that the Bitcoin protocol can be used to implement a data store with CRUD operations. Its {\UTXO} model offers better flexibility in database transaction formulations and better concurrent execution optimizations. \revise{This opens up new potential in system design that benefits from the decentralization of the blockchain without compromising too much on the performance.}

Therefore, this work proposes a new {\UTXO}-based approach to the combination of blockchain and database that takes both verifiability and decentralization into account. We aim to achieve decentralization in these aspects: (1) Data Ownership: The data owner solely relies on permissionless blockchains for ownership protection. (2) Data Access: Data operations authorized by the data owner can be directly processed by permissionless blockchains and never eclipsed by the database administrator. (3) Data Verification: Any data manipulations can be verified with proofs coming directly from the permissionless blockchain. (4) Data \revise{Provenance: The data history is immutable and publicly verifiable without relying on the honesty of a database administrator.} To sum up these goals, our vision is a database system where mutual trust between any pair of entities can be easily established using a blockchain. This type of truly decentralized data management paradigm can open up new potentials in systems such as supply chains~\cite{gurtu2019potential, saberi2019blockchain}, asset management~\cite {truong2023blockchain, zakhary2019towards, baltais2024economic}, IoTs~\cite{reyna2018blockchain, Maha2019BCIOT}, and healthcare~\cite{agbo2019blockchain, zhang2018blockchain}. \revise{Since the participating parties in these systems need to establish mutual trust over the "what" and "when" of data, a decentralized database system provides a new source of trust that is substantiated by a global-scale network. Compared to reputation-based trust, this is more quantifiable, cheaply verifiable, and suitable for a future agentic economy~\cite{hadfield2025economy}.}

To achieve these goals, we propose SpendableStore as the first {\UTXO}-based decentralized database that has been implemented and tested on a stably running permissionless blockchain. Our contribution includes: (1) We demonstrate how {\UTXO}s can be used to store arbitrary data with access control and perform CRUD operations beyond merely storing crypto coins. (2) We demonstrate how {\UTXO} blockchain transactions can be utilized to introduce transactional support to blockchain-based decentralized database systems. (3) We propose Future Now Snapshot Isolation as a blockchain-tailored variant of the snapshot isolation guarantee to provide transactional correctness when executing transactions in decentralized database systems. (4) We conduct experiments on permissionless blockchains to show that SpendableStore outperforms the existing data decentralization solutions in terms of transaction processing throughput and cost while providing a higher level of decentralization. 

The paper is organized as follows. Section~\ref{section: Background} provides the background information on {\UTXO} blockchains and a state-of-the-art blockchain-based database to highlight the limitations that will be addressed by our work. Section~\ref{section: SpendableDO} introduces Spendable Data Object as the atomic data unit in the SpendableStore design. Section~\ref{section: SpendableStore} describes the architecture and design details of SpendableStore and elaborates on transaction processing. Section~\ref{section: FNSI} explains the Future Now Snapshot Isolation and provides anomaly analysis. Section~\ref{section: Experiment} describes the implementation details and conducts experiments. Section~\ref{section: Related Works} discusses related works. Section~\ref{section: Conclusion} concludes.

\section{Background}
\label{section: Background}
This section provides information about {\UTXO} blockchains, transaction processing, and blockchain-based databases.

\subsection{{\UTXO} Blockchain Overview}
\label{subsection:bc_overview}

\begin{figure}
    \centering
    \includegraphics[width=\linewidth]{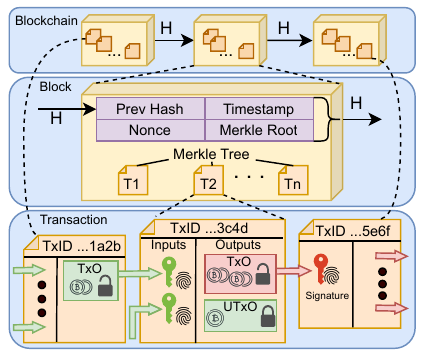}
    \caption{Blockchain Structure.}
    \vspace{-1.5em}
    \label{fig:blockchain_structure}
\end{figure}

\textbf{Blockchain.}
A blockchain is a series of blocks of transaction records connected by cryptographic hashes. Each block consists of a header and a transaction payload structured as a Merkle Tree. The header contains metadata information such as the previous block's hash and the Merkle root. The top part of Fig.~\ref{fig:blockchain_structure} visualizes three consecutive blocks of a blockchain and the internals of a block. The letter \textit{H} represents the hash value of the block header. 

\textbf{Decentralized Protocol.}
Permissionless blockchains are networks running under an open protocol that defines the validation of transactions and blocks and a consensus mechanism. 
A full node maintains the full blockchain copy and can provide query services. A mining node verifies new transactions and creates new blocks. A Simplified Payment Verification (SPV) Node only maintains the blockchain headers to provide quick transaction validation~\cite{nakamoto2008bitcoin}. 

\textbf{{\UTXO}.} 
A {\UTXO} transaction, as illustrated in the bottom segment of Fig.~\ref{fig:blockchain_structure}, has a unique transaction ID, a list of inputs, and a list of outputs. It consumes the inputs and produces the outputs. Each produced transaction output {\TXO} is uniquely identified by the combination of the transaction ID and its index in the output list. Each {\TXO} can be \textit{spent} as an input of a later transaction exactly once. {\TXO}s that have not been \textit{spent} are called Unspent Transaction Outputs ({\UTXO}) and hold coins. Once \textit{spent}, the coins are transferred into the new {\TXO}s. {\UTXO} transactions form a Directed Acyclic Graph.

Each {\UTXO} contains a locking script that determines how it can be \textit{spent}. A transaction attempting to \textit{spend} it must provide a \textit{unlocking script} as if providing a key to a lock. The \textit{unlocking script} could be a digital signature, the preimage of a hash value, or something more complex, depending on the \textit{locking script}'s code logic. \revise{Our work relies on a unique \textit{locking script} design called PUSHTX~\cite{zhang2021pushtx} which enables stateful contracts on {\UTXO} blockchains. At a high level, these script designs instruct the blockchain nodes to check if an output fulfills predefined rules, such as requiring that certain data be carried over from the input. } 

In Fig.~\ref{fig:blockchain_structure}, transaction 3c4d sends some coins from Alice (green) to Bob (red). It has two inputs, one comes from transaction 1a2b in a previous block, and the other is from another unshown transaction.
Alice unlocks both inputs with her digital signature and specifies two outputs. The first contains the desired amount of coins for Bob and a locking script that can only be unlocked by Bob's digital signature. The second contains the excessive coins and can be unlocked only by Alice herself. In a later transaction 5e6f, Bob spends the received coins by unlocking the {\UTXO} with his signature. 

\subsection{{\UTXO} Transaction Processing}
\label{subsection:bc_transaction_processing}

\textbf{Construction.}
Each transaction is first fully constructed with the input list, the corresponding \textit{unlocking script}s, and the output list arranged into a byte string according to the protocol. This step can be completed entirely off-chain as long as the inputs are locally cached.
Off-chain construction guarantees an atomic and deterministic execution outcome. Since the outputs are fully specified in this step, either all outputs become valid {\UTXO}s or none do, depending on the final status of the transaction.

\textbf{Validation.} 
Constructed transactions are sent to the mining nodes to be validated. Transactions that attempt to \textit{spend} a non-existing {\UTXO}, double-\textit{spend} a {\UTXO}, or provide a wrong \textit{unlocking script} are invalidated. Each mining node independently validates all received transactions based on its current known {\UTXO} set. The invalid ones are discarded while the valid ones are added to a local buffer, the \textit{Memory Pool} (Mempool). The validation result is immediately returned to the transaction sender. The validation time cost is in the magnitude of milliseconds, while message propagation adds additional delay. Overall, the sender typically receives the validation results in less than a second.



\textbf{Confirmation.} 
A transaction receives its first confirmation when it is included in a new block. It gains one more confirmation for each consecutive block in the future. A confirmation indicates that the network has reached a consensus on the partial ordering of all transactions and total orderings of those that are related by the \textit{spending} relation, up to the block. A higher number of confirmations exponentially increases the probability that this consensus will remain permanent~\cite{nakamoto2008bitcoin}. 

\subsection{Blockchain-based Databases}
\label{subsection:data_on_bc}
We discuss a representative example of blockchain-based database---BlockchainDB~\cite{Hindi2019BlockchainDB}---to highlight its limitations.

BlockchainDB uses Ethereum Smart Contracts to maintain data tables in the memory of the Ethereum Virtual Machine and exposes a function to modify a single specified row in the table. It deploys multiple contract instances for data sharding to achieve better performance. BlockchainDB has the following characteristics and drawbacks that we see as areas of potential improvement.

(1) Poor support for data ownership. In BlockchainDB, each smart contract can only be invoked by a dedicated server node called a BlockchainDB Peer. The peers form a database layer that separates clients from the blockchain. All client requests must go to the database layer and be converted to respective blockchain function calls. 

(2) Inability to process complex transactions. BlockchainDB does not support database transactions that update multiple data items simultaneously. Its smart contracts can only update a single row of the table in each function call; therefore, a client request that updates two rows will be converted into two blockchain transactions that are independently validated. If only one is validated, the client will observe partial effects. 

(3) Poor support for decentralization. BlockchainDB was deployed on a private blockchain, which can be configured to allow data sharding. Its experiment settings and configurations cannot be replicated in a permissionless blockchain because the protocol cannot be modified, thus not supporting sharding. Its experiment results show that its transaction throughput is at most 25 updates per second if deployed to the current Ethereum blockchain.

\section{Spendable Data Object}
\label{section: SpendableDO}
In this section, we define Spendable Data Object ({\SpendableDO}) as the atomic data unit of {\SpendableStore}.

\subsection{Overview}
\label{subsection: SDO_overview}
A data object in {\SpendableStore} is represented by a series of blockchain transaction outputs ({\TXO}), each encapsulating a snapshot of the data object. Two consecutive versions are connected by a blockchain transaction, which \textit{spends} the previous version {\TXO} and outputs a {\TXO} for the next version. Due to this structural similarity to \textit{spending} crypto coins, we refer to a data object in SpendableStore as a Spendable Data Object ({\SpendableDO}). And we refer to a {\TXO} version of a {\SpendableDO} as an Output Version or, equivalently, an Object Version ({\oVersion}).


The {\oVersion}s of a {\SpendableDO} always form a singly-linked append-only list on the blockchain, which manifests the complete lineage of the {\SpendableDO}. This is because each {\oVersion} can be \textit{spent} only once, and the transaction spending it always outputs the successive {\oVersion}. The current state of the {\SpendableDO} is the {\oVersion} at the end of the list; it is the only {\oVersion} that has not been \textit{spent} yet. Changing the state of a {\SpendableDO} is done by a transaction that \textit{spends} this {\oVersion} and outputs the new current state. \revise{ The state in a {\SpendableDO} is a byte sequence that can be used to represent arbitrary data structures. In the rest of the section, we use key-value pairs (KVP) as a representative type of {\SpendableDO} because their duality allows for a straightforward illustration of role-based access control in {\SpendableDO}s, which is critical for a decentralized database. } 

\begin{figure}
    \centering
    \includegraphics[width=\linewidth]{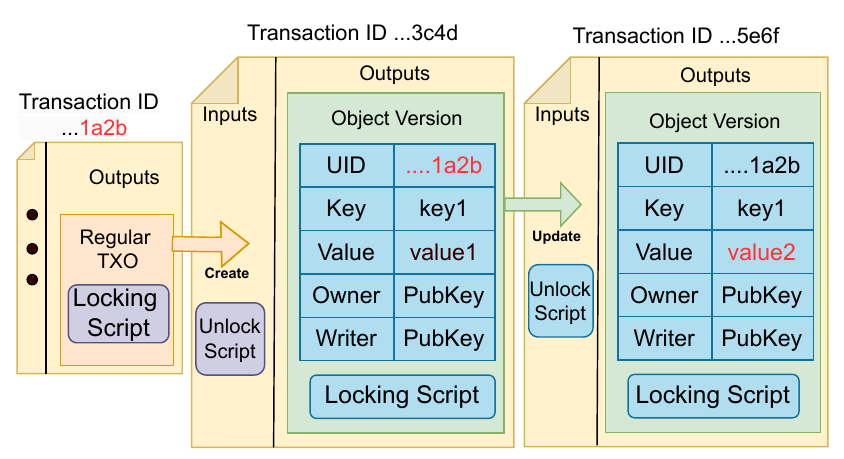}
    \caption{Create and Update a SpendableDO.}
    \vspace{-1.5em}
    \label{fig: SDO-CRUD}
\end{figure}

The green areas in Fig.~\ref{fig: SDO-CRUD} show a simplified visualization of a {\SpendableDO}. It shows two {\oVersion}s of a KVP with a Unique ID ending with 1a2b. The left one is the older {\oVersion} with value "value1" and the new {\oVersion} changes the value while keeping all other fields identical. Transaction 5e6f manifests an update-value operation on KVP "1a2b". It \textit{spends} the old {\oVersion} with a correct unlocking script and outputs the updated {\oVersion}. 

\subsection{Content}
\label{subsection: SDO_content}
Each {\oVersion} is embedded with the following list of data fields:

\begin{description}
    \setlength{\itemsep}{0pt} 
    \setlength{\topsep}{0pt} 
    \item[UID] A universally unique identification number given to the contained KVP at the time of its creation. 
    \item[Key] A sequence of bytes to represent the KVP key.
    \item[Value] A sequence of bytes to represent the KVP value.
    \item[Owner] The public key of the KVP owner.
    \item[Writer] The public key of the entity that was granted write access privileges of the KVP.
\end{description}

The UID uniquely identifies a KVP in the entire blockchain network and remains consistent throughout its lifecycle. The choice of UID is related to the {\SpendableDO}'s creation operation. We provide more details about this process in Section~\ref{subsection: SDO_operations}. 
The key and value need to be encoded into byte strings, which are upper-bounded by the blockchain protocol at around 4GB~\cite{PushdataOpcodes}. The remaining two fields are used to enforce ownership and facilitate access control when validating the unlocking scripts. The owner has the privilege to modify all data fields except for the UID, while the writer can only modify the value field. 

The locking script is not part of the data fields but is a critical component that determines how to change the state of the KVP. It can only be unlocked by a transaction if all the following conditions are met. (1) The transaction outputs an {\oVersion}. (2) The outputted {\oVersion} is the result of applying a legal {\SpendableDO} operation. (3) The transaction is digitally signed by an entity that has the privilege to invoke the operation applied. \revise{We enforce these conditions by using the PUSHTX~\cite{zhang2021pushtx} method in our locking scripts.}

 
\subsection{CRUD Operations}
\label{subsection: SDO_operations}
\textbf{Create.}
A new KVP is created by a blockchain transaction that outputs the initial state {\oVersion}, illustrated in Fig.~\ref{fig: SDO-CRUD}, in which transaction 3c4d in the middle creates a KVP with a UID ending with 1a2b. This transaction does not take an {\oVersion} as input but rather a regular {\TXO} that carries crypto coins. This {\TXO} pays the transaction fee and determines the KVP's UID. Since each transaction ID is universally unique, we took advantage of this by borrowing this ID as the KVP's UID, highlighted in red. This guarantees UID uniqueness because each {\TXO} can only be \textit{spent} once to create a single KVP. 

The "create" transaction goes through the construction, validation, and confirmation steps like other transactions. An off-chain entity constructs the transaction by initializing the data fields and programming a corresponding locking script. 

\textbf{Read.}
Reading a KVP means to read a specific {\oVersion} of the KVP. Since each {\oVersion} can be found in the output list of a specific transaction, anyone can query any full node for the transaction raw bytes as long as they provide the transaction ID. They can then deserialize the returned bytes to obtain the KVP state. They can also request the corresponding Merkle proof and verify independently with an SPV node. The read operations do not involve the construction of new transactions and are not recorded on the blockchain. 

Reading any {\oVersion} reveals the origin of the KVP because the UID is the transaction ID that precedes the "create" transaction. They can also recursively trace their previous {\oVersion}s based on the information in the input list. Some full nodes might maintain additional indexes that remember which transaction \textit{spent} a given {\TXO}. These full nodes can be queried to get the successive {\oVersion}s, completing the full data history. 

\textbf{Update.}
Each KVP can be updated in four ways: (1) The owner can modify the Key and Value fields. (2) The writer can modify the Value field only. (3) The owner can modify the Writer field to grant or revoke writing privileges. (4) The owner can modify the Owner field to transfer ownership. 

Every update on a KVP is reflected by a transaction recorded on the blockchain. The transaction specifies which update method was applied in its unlocking script, and the output {\oVersion} must be consistent with that method. This means that only the relevant data fields can be changed. Additionally, the unlocking script must also include the digital signature of the changes by the entity with the correct privileges. If the transaction fails to meet these requirements, it will not be validated by the mining nodes.

\textbf{Delete.}
KVPs' history records are all immutable {\oVersion}s on the blockchain; therefore, we employ tombstone deletion. One way is to update the Key and value both to null, as this will make the {\SpendableDO} appear deleted in a higher-level data structure. Another way is to update both the Owner and Writer to null. Doing so prevents any entity from updating the {\SpendableDO} permanently because no one would be able to generate a valid signature for it anymore. This approach is irreversible because resetting an owner requires an update operation. 

\subsection{Atomic Multi-Updates}
\label{subsection: SDO_multi_updates}
Multiple KVPs can be updated atomically in a single blockchain transaction. Since {\UTXO} transactions can take multiple inputs and outputs, one can construct such a transaction where each input-output pair corresponds to one KVP. Each involved KVP can be applied with a different update method, but their validation status is now atomic because any single invalid KVP update nullifies the entire transaction.

The input {\oVersion}s in a multi-update transaction may not have the same Writer or Owner. Two entities can collaboratively construct the transaction and provide digital signatures individually over the entire transaction. The input {\oVersion}s may have different internal structures and represent different data types. For example, an {\oVersion} for a KVP and an {\oVersion} for a graph node can be updated atomically in one transaction. This is a new approach to solving the cross-database transaction problem as long as the involved databases are all {\SpendableDO}-based. Moreover, the transaction can include crypto-coin inputs and outputs for monetary transfers, allowing for more customizable database designs that involve payments.

\subsection{Properties}
\label{subsection: SDO_properties}
To summarize, the following properties of {\SpendableDO} make it suitable for a decentralized database:

\textbf{Availability and Durability.} 
As part of blockchain transactions, {\SpendableDO}s are fully replicated across the blockchain network. Because of the protocol's built-in transaction fee mechanism, all network nodes are properly incentivized to provide storage service for all transactions. Therefore, {\SpendableDO} operations can be handled by any node in a worldwide distributed network and always produce consistent verifiable responses. This reduces the risk of a compromised central storage taking data hostage. 

\textbf{Fine-grained Ownership.}
{\SpendableDO} has a data-object level ownership protection and access control that are enforced collectively by the blockchain. Unlike centralized database designs, {\SpendableDO} owners do not need to trust a central database administrator for data security. Additionally, any permission changes are transparent, verifiable, and immediately effective. 

\textbf{Deterministic Outcome.}
Update operations on {\SpendableDO}s are given to the decentralized network as transactions with known deterministic outputs. This benefits the data owners because they do not need to worry that an unanticipated change will be applied. It also reduces the workload on mining nodes since they only perform verification instead of computation~\cite{eberhardt2017or}. 

\section{Spendable Store}
\label{section: SpendableStore}
This section describes our design of {\SpendableStore} as a fully decentralized blockchain-based data store that can be used as the storage and transaction management layers of future database designs. 

\begin{figure}[ht]
    \centering
    \includegraphics[width=0.8\linewidth]{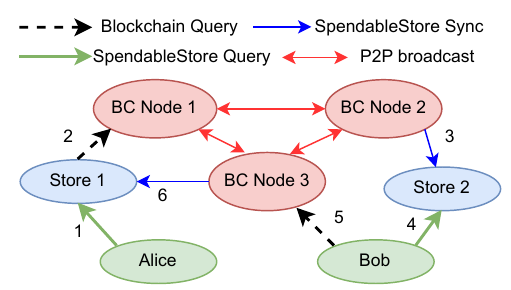}
    \caption{SpendableStore System Model.}
    \vspace{-1em}
    \label{fig: SpendableStore-Architect}
\end{figure} 
\vspace{-1em}

\subsection{System Model}
\label{subsection: SpendableStore_System_Model}
Three types of entities participate in the {\SpendableStore} system: the blockchain, the SpendableStore instances~({\SSInstance}), and the data owners. The blockchain passively serves as the storage, backup, and access control layer. The {\SSInstance}s act as the query processing, optimization, and transaction management layer. The data owners are like traditional database clients except that they can bypass the {\SSInstance}s to interact with their data in the storage layer directly. 

Fig.~\ref{fig: SpendableStore-Architect} provides a system diagram with three blockchain nodes, two {\SSInstance}s, and two data owners. It shows the four types of interactions with colored links. They include the blockchain nodes' internal communication (red); blockchain queries from {\SSInstance}s and data owners to the blockchain (black); synchronization connections from {\SSInstance}s to blockchain nodes (blue); and {\SpendableStore} queries sent by data owners to a {\SSInstance} (green).

\textbf{Blockchain.}
Each blockchain node physically stores the {\SpendableDO}s and processes all their operations through the validation and confirmation steps. They also resolve conflicting operations through the blockchain consensus mechanism to yield a single version of history for every {\SpendableDO}. We explained this in more detail in Sections~\ref{subsection:bc_transaction_processing} and ~\ref{subsection: SDO_operations}. 

The blockchain nodes provide two basic query windows. (1)~getTransaction: This function takes a transaction ID and returns the requested transaction if it exists. The response includes the inputs, the outputs, and metadata such as the Merkle proof. (2)~broadcastTransaction: This function takes a fully constructed transaction to be included in the blockchain. If the transaction is validated, its transaction ID will be returned. Otherwise, it returns an error message. 

\textbf{SpendableStore instances.}
A {\SSInstance} manages a subset of the {\SpendableDO}s on the blockchain and provides a database view to the data owners. It maintains high-level data structures like indexes, tables, and graphs over the {\SpendableDO}s based on what they represent. It provides the complementary query interface to the data owners, which are collectively called {\SpendableStore} queries. Handling these queries requires a {\SSInstance} to be synchronized with the blockchain and to construct the blockchain transaction to carry out {\SpendableDO} operations. Therefore, it maintains persistent connections with the blockchain nodes and sends blockchain queries such as "boradcastTransaction". We elaborate on this in Section~\ref{subsection: Instance_Design}

Multiple independent and autonomous {\SSInstance}s can exist on the same blockchain. Each {\SSInstance} securely possesses a private key and is uniquely identified by the corresponding public key. The sets of {\SpendableDO}s managed by two {\SSInstance}s can overlap partially or fully. This happens if the overlapped {\SpendableDO} is relevant to two databases or if one {\SSInstance} is the replica of another. A {\SSInstance} has the write privilege to a {\SpendableDO} if its public key matches the Writer field. Since the field can contain one key only, at most one {\SSInstance} can write to a {\SpendableDO} at a time. {\SSInstance}s do not need any form of communication to resolve conflicts and may not even know about the other's existence. This is possible because all {\SSInstance}s refer to the same blockchain for the ground truth on any {\SpendableDO}s.

\textbf{Data owners.}
A data owner securely possesses a private key and generates valid signatures independently. It owns a {\SpendableDO} if its public key matches the Owner field. 
A data owner can directly interact with the blockchain through blockchain queries to perform {\SpendableDO} operations. A data owner grants write privileges to a {\SSInstance} through the "NameWriter" update operation. It can then send {\SpendableStore} queries to interact with her {\SpendableDO}s. It can revoke or redistribute the write privileges at any time. A data owner runs an SPV node. It requires a minimum amount of hardware~\cite{nakamoto2008bitcoin} and allows the data owner to verify the responses of the blockchain queries and {\SpendableStore} queries. 

\textbf{Example.}
Assume Fig.~\ref{fig: SpendableStore-Architect} represents a global supply chain system using {\SpendableStore} \revise{where the supply chain items are digitized into real-world asset tokens (RWA)~\cite{baltais2024economic} as individual {\SpendableDO}s.} The two {\SSInstance}s (blue ellipses) are located in two regions, each maintaining the {\SpendableDO} tokens of the items in the regional factory. Alice now transfers some items to Bob, and it involves the state-changes of certain {\SpendableDO}s \revise{such as changing custody and receiving a quality control certificate}. The process is shown by arrows labeled 1 to 6. (1) Alice initiates the changes by sending {\SpendableStore} queries to {\SSInstance} 1. 
(2) The queries are converted into blockchain transactions and sent to BC node 1. The blockchain internally broadcasts and verifies the transactions. (3) {\SSInstance} 2 synchronizes with BC node 2 and updates its local copy. (4) Bob verifies the completion of the transfer by sending {\SpendableStore} read queries to {\SSInstance} 2. (5) Bob learns more about this supply chain item's history and performs additional update operations by sending blockchain queries directly to BC node 3. (6) {\SSInstance} 1 observes Bob's updates through synchronization. If Bob's actions are still relevant to {\SSInstance} 1, it can capture them through blockchain synchronization. Otherwise, {\SSInstance} 1 can remove all involved {\SpendableDO}s' locally cached copies without worrying about permanent data loss. 

\subsection{Security and Safety Models}
\label{subsection: SpendableStore_Safety_Model}

\removed{Data owners' trust in a {\SSInstance} depends on the entity maintaining it. An anonymous {\SSInstance} should be avoided, while those run by well-known organizations can be trusted more. However, even the most secure organization might be temporarily compromised to act maliciously. }

{\SpendableStore}'s design ensures that the extent of damage a malicious {\SSInstance} can cause is constrained. Assuming data owners' private keys are securely protected, the following safety guarantees are provided: (1) {\SpendableDO} ownership safety: malicious {\SSInstance}s can never alter the owner of any {\SpendableDO}s. (2) Detectable tampering: the history states of a {\SpendableDO} are immutable. A malicious {\SSInstance} without the Writer privilege of a {\SpendableDO} cannot tamper with the current state through updates. Compromised writers may tamper with the state using a public blockchain transaction, but this is easily detectable. The data owner can revoke an {\SSInstance}'s writer privilege by the NameWriter operation. (3) No single point of failure: If a {\SSInstance} fails, the data owners can switch to another {\SSInstance} with low overhead. (4) Verifiable query responses: All {\SpendableStore} queries can be verified by querying a blockchain node for the related blockchain transactions. \revise{Because of these guarantees, the added layer of {\SpendableStore} instances does not undermine the decentralized nature of the blockchain. The data owners always have the freedom to bypass this layer, as illustrated in Fig.~\ref{fig: SpendableStore-Architect}.}

\subsection{Instance Design}
\label{subsection: Instance_Design}
We provide the {\SSInstance} design for a decentralized Key-Value store based on the KVP {\SpendableDO} from Section~\ref{section: SpendableDO}. It has three main functionalities. First, it stores all {\oVersion}s of a set of KVPs locally as a partial cache for the blockchain. Second, it maintains indexes for efficient lookups and searches over the KVPs. Lastly, it handles {\SpendableStore} Queries from Data owners by constructing {\UTXO}-transactions to perform {\SpendableDO} operations on the blockchain. 

\textbf{Query Interface.}
The {\SSInstance} provides a customized query interface based on the types and semantics of the {\SpendableDO}s. For example, a simple Key-Value Store may have the following simplistic interface:

\begin{description}[itemsep=0pt]
    \item [ListAll](in: null; out: all UIDs)
    \item [InsertKVP](in: PubKey, key, value; out: TxID, UID)
    \item [GetKVP](in: UID; out: KVP)
    \item [UpdateKVP](in: UID, value; out: TxID)
    \item [DeleteKVP](in: UID; out: TxID)
\end{description}

\textit{ListAll} returns the list of the UIDs of all KVPs currently in the store. \textit{InsertKVP} takes a public key, a key string, and a value string as inputs and performs the "create" operation. The created {\SpendableDO}'s Owner field is initialized to the input public key, and the Writer field is initialized to the {\SSInstance}'s public key. It returns the blockchain transaction ID and the KVP's assigned UID. \textit{GetKVP}, \textit{UpdateKVP}, and \textit{DeleteKVP} correspond to a {\SpendableDO}'s read, update, and delete operations.

The {\SSInstance} may additionally provide interfaces depending on the application it serves. Such as a binary search tree formed using {\SpendableDO}s as tree nodes.

\begin{figure}
    \centering
    \includegraphics[width=0.8\linewidth]{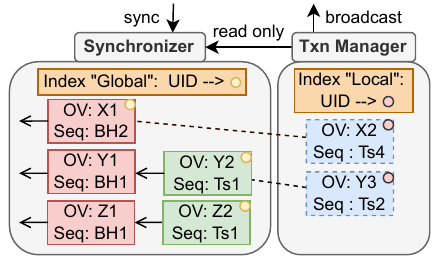}
    \caption{SpendableStore Instance Internal Structure.}
    \vspace{-1.5em}
    \label{fig: SpendableStore-Internal}
\end{figure}

\textbf{Internal Structure.}
The {\SSInstance} has two major components: a blockchain synchronizer and a transaction manager (Fig.~\ref{fig: SpendableStore-Internal}). The synchronizer keeps track of the latest global states of all the stored KVPs. It subscribes to the blockchain for any newly validated or confirmed transaction that performs a {\SpendableDO} operation. These transactions might be initiated by the {\SSInstance} itself, another {\SSInstance}, or individual data owners. They indicate changes in the global state of the KVP store, so the synchronizer updates the local state accordingly.

The synchronizer manages the global states of each KVP in a linked list where each node contains an {\oVersion} and a sequence number. A newly received mempool-validated {\oVersion} is linked after its previous {\oVersion}, and the sequence number is set to the timestamp of receiving. When the synchronizer later learns that the {\oVersion} is confirmed in a block, it updates the sequence number to the height of the block. The sequence numbers are used for concurrency control and are explained in later sections. The synchronizer also maintains a hash index "Global" for quick access to the most recent known global states. It maps a KVP's UID to a pointer that points to the head node of the KVP's linked list. 

The left half of Fig.~\ref{fig: SpendableStore-Internal} shows a synchronizer that manages the blockchain-confirmed {\oVersion}s (red) and mempool-validated {\oVersion}s (green) of three KVPs X, Y, and Z. The latest global {\oVersion} for each KVP is marked by a yellow circle in the upper-right corner. X's most recent {\oVersion}, X1, was confirmed in the block with height two (BH2). Y and Z's most recent {\oVersion}s are still in the Mempool, and the synchronizer learned about them at local time Ts1. 

The transaction manager handles incoming SpendableStore queries based on the KVP store's states maintained by the synchronizer. For read-only queries, it reads from the synchronizer and directly returns the relevant {\oVersion}s. Otherwise, it additionally builds blockchain transactions to perform a relevant {\SpendableDO} operation to produce new {\oVersion}(s). It sends the transaction to the blockchain via the \textit{broadcastTransaction} query and asynchronously waits for the validation response. 

While waiting for the validation, it temporarily stores the newly produced {\oVersion}s in linked lists similar to those in the synchronizer. It maintains a hash index "Local" for quick access to these local states. We further explain how the local states are used to improve concurrency in Section~\ref{section: FNSI}.

The right half of Fig.~\ref{fig: SpendableStore-Internal} shows the complementary transaction manager. It is processing some queries that update X and Y. It has produced two new local {\oVersion}s, X2 and Y3 (blue), and is waiting for their validation results. Once validated, the transaction manager removes these {\oVersion} from the "Local" index, while the synchronizer adds them to the "Global" index. 

\textbf{SpendableStore-Transactions.}
The transaction manager can process multiple queries in a transaction to provide transactional properties. We formally define such an atomic set of queries as a {\SpendableStore}-transaction ({\SSTx}) to differentiate it from {\UTXO}-blockchain transactions. If an {\SSTx} is not read-only, a blockchain transaction must be constructed to carry out the corresponding {\SpendableDO} operation(s). We formally define this blockchain transaction as the Blockchain Twin (BC-Twin) of the {\SSTx}. For an {\SSTx} that updates multiple KVPs, its BC-Twin will be a multi-update blockchain transaction as specified earlier in Section~\ref{subsection: SDO_multi_updates}. 

\begin{figure}
    \includegraphics[width=0.4\textwidth]{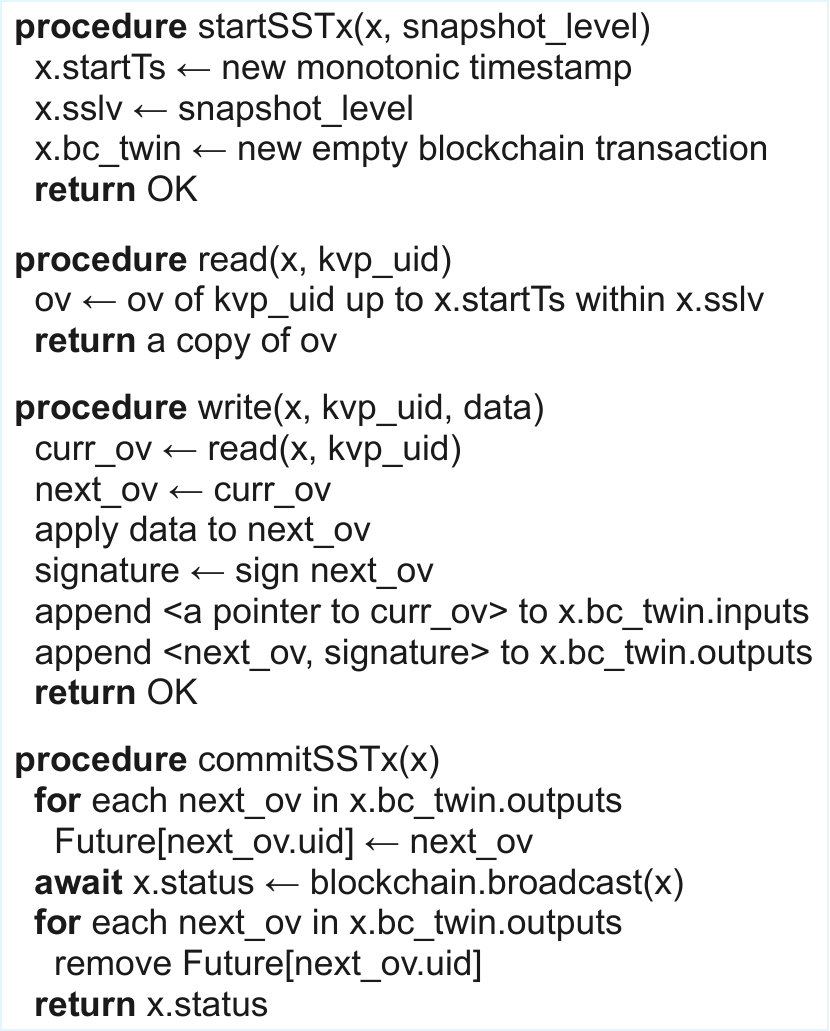}
    \caption{{\SSTx} Processing.}
    \vspace{-1em}    
    \label{fig: SS-Tx_psudocode}
\end{figure}

Fig.~\ref{fig: SS-Tx_psudocode} gives the pseudocode of the four procedures for processing an {\SSTx}. The first procedure, "startSSTx", assigns a start timestamp, sets a specified snapshot level, and initializes an empty BC-Twin. The snapshot level determines what {\oVersion}s are visible to the {\SSTx}; we elaborate on this in Section~\ref{section: FNSI}. The "read" procedure returns a copy of the most recent {\oVersion} of a KVP as of the start timestamp within the range of the snapshot level. The "write" procedure always performs the "read" first to obtain the current {\oVersion}, then creates a {\oVersion} with the new data. The two {\oVersion}s are added as an input/output pair to the BC-Twin. Since the BC-Twin is only modified in the "write" but not in the "read", it carries no information about the read-only KVPs in the original {\SSTx}. This is not a limitation of BC-Twin but a performance optimization choice which we will explain in Section~\ref{section: FNSI}. The last procedure "commitSSTx" writes the new {\oVersion}s temporarily into the "Local" index in the Txn Manager's memory. Then it sends the BC-Twin to the blockchain to be validated. Once any validation result is returned, the {\oVersion}s in the "Local" index are removed. 

\section{Future Now Snapshot Isolation}
\label{section: FNSI}
In this section, we propose Future-Now Snapshot Isolation ({\FNSI}) as a snapshot isolation guarantee for decentralized databases like {\SpendableStore}. We formally define it (Section~\ref{subsection: FNSI_Overview}), then analyze its unique characteristics and the anomalies it prevents (Section~\ref{subsection: FNSI_analysis}). 

\subsection{FNSI Overview}
\label{subsection: FNSI_Overview}
We propose FNSI as a new snapshot isolation protocol adapted for decentralized database transactions processed through {\UTXO}-based blockchains, i.e., {\SSTx}s. It is a relaxation of Snapshot Isolation (SI) guarantees. Achieving Serializability (SR) or SI in {\SpendableStore} would incur prohibitively large computation and communication overheads that render the system impractical. On the other hand, FNSI can prevent most transactional anomalies, such as dirty reads and non-repeatable reads, at a low additional overhead. 

A naive technique to force a globally serializable execution of {\SSTx}s is to \textit{spend} all the reads in addition to the writes. This means to include all the {\oVersion}s read but not written by an {\SSTx} also as inputs to its BC-Twin and make the BC-Twin outputs unmodified {\oVersion}s. Doing so implies that any two {\SSTx}s with any type of conflict---write-write, read-write, or write-read---are guaranteed to be ordered on the blockchain. However, this technique inevitably prohibits concurrent reads, since read-read conflicts are ordered as well, and thus leads to a significant performance drop. It also easily bloats up the BC-Twin's size and the transaction fee cost. Although it can be used occasionally for {\SSTx}s that require SR, it is not suitable for the general operation of {\SpendableStore}.

SI is often used as a relaxed alternative to SR. However, as earlier works~\cite{Sovran2011PSI, elnikety2004generalized} demonstrated, SI cannot be directly used in replicated databases with asynchronous communications since it is infeasible to obtain a globally consistent snapshot. In {\SpendableStore}, synchronization among all {\SSInstance}s is only passively obtained as a byproduct of each new blockchain block. Between blocks, {\SSInstance}s may have different or conflicting views about the database state. Given the low block generation frequency, {\SpendableStore} cannot support a realistic database workload under SI. 

FNSI is a variation of SI that allows an {\SSInstance} to take a snapshot of the most recent database state to the best of its knowledge as of the start of an {\SSTx}. The snapshot may contain states that are not globally consistent or available as long as they are locally consistent. Specifically, a locally consistent snapshot must have a conflict-free {\oVersion} history for each known {\SpendableDO} and does not contain any partial effects of an {\SSTx}, i.e., an incomplete output list of a BC-Twin.

Unlike other SI protocols, where a snapshot only contains globally known committed transactions, FNSI allows a snapshot to additionally contain committed {\SSTx}s still in the network propagation. Therefore, a snapshot in FNSI may include the "Future", represented by the {\SSTx}s that are not recorded in the blockchain, on top of "Now", which are {\SSTx}s that are already included in the globally consistent blockchain; hence the name Future Now Snapshot Isolation. 

In practice, FNSI defines three snapshot levels based on how much of the "Future" is included in the snapshot.

\textbf{(1) Blockchain Snapshot (BCSS):} A BCSS is a global snapshot passively taken once per new block and purely consists of "Now". A BCSS is associated with a block height number and includes the state of the database recorded in the blockchain up to that block. In Fig.~\ref{fig: SpendableStore-Internal}, the most recent BCSS is of height two and consists of {\oVersion} X1, Y1, and Z1. 

\textbf{(2) Mempool Snapshot (MPSS):} An MPSS is an approximated global snapshot that can be taken at any time by a {\SSInstance} and consists of the "Now" and the validated "Future". An MPSS has a block height number and a local timestamp. It includes the most recent global state known by the local synchronizer as of the timestamp, either in the blockchain or in the mempool. In Fig.~\ref{fig: SpendableStore-Internal}, a MPSS as of timestamp Ts2 consists of {\oVersion} X1, Y2, and Z2.

\textbf{(3) Local Snapshot (LCSS):} A LCSS is a local snapshot that can be taken at any time by a {\SSInstance} and consists of the "Now" and all the "Future" known locally. An LCSS has a block height number and a local timestamp. It includes the most recent state accessible by the local transaction manager as of the timestamp, even if not validated yet. In Fig.~\ref{fig: SpendableStore-Internal}, the LCSS as of timestamp Ts4 consists of {\oVersion} X2, Y3, and Z2. 

A {\SSInstance} processes each {\SSTx} under one of the three snapshot levels chosen by the data owner. Whenever the {\SSTx} requests to read a {\SpendableDO}, the most recent {\oVersion} as of the start timestamp within the snapshot level will be returned. For example, in Fig.~\ref{fig: SpendableStore-Internal}, if an {\SSTx} chooses LCSS and is assigned a start timestamp Ts3, it reads {\oVersion}s X1, Y3, and Z2.

\subsection{FNSI Analysis}
\label{subsection: FNSI_analysis}

\textbf{Commitment.}
There are three levels of commitment for an {\SSTx}: "block committed", "instance committed", and "inferred committed", which correspond to the three levels of snapshots. An {\SSTx} is "block committed" if its BC-Twin is confirmed in a block. This is the highest level of commitment because the block is a verifiable consensus among the blockchains about the validity of the {\SSTx}. All participating entities in the {\SpendableStore} can independently verify the successful execution of the {\SSTx}.

An {\SSTx} is "instance committed" after its BC-Twin is validated. However, this commitment level is restricted to the {\SSInstance} that processes the {\SSTx}. Since the {\SSInstance} must have exclusive Writer privilege on all the input {\oVersion}s to construct a valid BC-Twin, it is sure that no other {\SSInstance} can create another valid but conflicting BC-Twin. Knowing that there would not be any potential conflicts, the {\SSInstance} is sure that the {\SSTx} will eventually be "block committed". \revise{Although a malicious data owner may collude with another SS-Instance, it must first transfer the writer privilege via an Update-Writer operation in a standalone transaction. This transaction does not replace the "instance committed" BC-Twin but follows it on the blockchain, exposing the malicious nature of the data owner.} 

An {\SSTx} can be considered "inferred committed" under a specific condition, even if its BC-Twins have not been validated. Assume a {\SSInstance} concurrently processes two {\SSTx}s, A and B, that update the same {\SpendableDO} X. A is currently inside its "commitSSTx" procedure on the await statement. B is set to the LCSS level, and it reads the {\oVersion} of X outputted by A, which A has written in the Future index. And B uses this {\oVersion} as an input in its BC-Twin. In this case, A is "inferred committed" from the perspective of B. If A does not end up in the blockchain for any reason, B will have an invalid input and thus not be part of the blockchain. In other words, A must be committed for B to be committed.

\begin{table}[t]
    \centering
    \caption{Concurrency Anomalies Analysis.}
    \label{table: anomalies}
    \begin{tabular}{|p{2.4cm}|c|c|c|c|c|}
        \hline 
        Anomaly & 
        SR &
        SI &
        \multicolumn{3}{|c|}{ \centering FNSI } \\
        \hline
        \hline
        {}                    &       &       & BCSS & MPSS & LCSS \\
        Dirty read            &  No   & No    & No    & No*   & Yes* \\
        Non-repeatable read   &  No   & No    & No    & No*   & Yes* \\
        Lost update           &  No   & No    & No    & No    & No   \\
        Write skew            &  No   & Yes   & Yes   & Yes   & Yes  \\
        \hline
        Long fork             &  No   & No    & No    & Yes   & Yes  \\
        Conflicting fork      &  No   & No    & No    & No    & No   \\
        \hline
    \end{tabular}
    \vspace{-1em}
\end{table}

\textbf{Anomalies.}
Table~\ref{table: anomalies} summarizes the anomalies allowed by different concurrency levels. The last two anomalies are uniquely present in database systems with asynchronous communications and have been defined and studied in ~\cite{Sovran2011PSI}. They refer to the division of database state and should be distinguished from blockchain forks. A long fork happens when two {\SSTx}s are simultaneously processed based on two nonidentical snapshots. A long fork must always merge back, without any application-specific or ad-hoc interference, such that all {\SSTx}s from all branches can be found in a unified conflict-free snapshot later in time. Otherwise, it is a conflicting fork. Some cells are asterisked to indicate that the anomaly analysis applies to general {\SSTx}s but can vary under certain scenarios. 

Serializability is the most desirable concurrency level; however, it is expensive for large workloads, as we have already mentioned. The snapshot-based concurrency levels all allow write skews to happen because the read-write conflicts among transactions isolated by snapshots are ignored. Except for write skews, SI provides the same level of guarantees as SR. 

The BCSS level in FNSI provides the same guarantees as SI. BCSS is not suitable for write-intensive workloads since it is infrequently updated and is often stale. 
However, BCSS is good for read-only {\SSTx}s and read-intensive workloads. The {\oVersion}s read under BCSS are all "block committed" with verifiable Merkle proofs and thus can be trusted even if they have been produced by unknown {\SSInstance}s and Data Owners. It is most recommended for OLAP workloads. 

The MPSS level is suitable for general workloads. It provides similar guarantees as SI, except for allowing long forks. Long forks happen under MPSS because each {\SSInstance} can have different views of the blockchain mempool, which results in nonidentical and conflicting snapshots across {\SSInstance}s. 
However, the long forks under MPSS are merged by each new block. The new block is a new global consistency checkpoint that resolves all write-write conflicts in the long forks. 

The MPSS level prevents dirty reads when an {\SSTx} always reads {\oVersion} that are "instance committed". In other words, the {\SSTx} is not reading a Mempool-{\oVersion} of a {\SpendableDO} whose Writer is another {\SSInstance}. Reading such {\oVersion}s has the potential danger of "read uncommitted" because the Writer {\SSInstance} might have sent out a conflict BC-Twins to different mining nodes of the blockchain and fabricated a temporary disagreement in the network. Likewise, reading "instance committed" is also necessary to prevent non-repeatable reads. 

Most database workloads can be processed under the requirement that all reads must be "instance committed". This is because database transactions often read and write to data objects that are semantically related; it is a reasonable assumption that these data objects have the same {\SSInstance} as the Writer. Additionally, if two {\SSInstance}s know each other and have sufficient trust over each other, they can treat each other's Mempool-{\oVersion} as "instance committed" to allow for smoother processing over inter-instance {\SSTx}s.

Lastly, LCSS is generally unsafe because the local-{\oVersion}s can be invalidated by the mining nodes. This happens even if the {\SSInstance} has the exclusive Writer privileges because the Query Engine processes incoming {\SSTx}s concurrently and might produce conflicting BC-Twins. Therefore, it may lead to dirty reads and non-repeatable reads. However, it is suitable for {\SSTx}s that write every {\SpendableDO} they read, i.e., with identical read and write sets, because of the "inferred commitment" rule. Since an earlier {\SSTx} is "inferred committed" from the perspective of a later {\SSTx}, it is no longer "read uncommitted" for the later {\SSTx} to read any outputted {\oVersion} of the earlier one, even if it has not been validated. LCSS provides the highest level of concurrency since the local snapshots are the least stale, thus minimizing write-write conflicts. It is the best option for OLTP workloads that read and write a set of {\SpendableDO}s together at high frequencies. 

Under all three isolation levels of FNSI, the "lost update" anomaly is always prevented. This is an intrinsic property of the {\UTXO}-blockchain transactions due to having deterministic outputs. Similarly, the "conflicting fork" anomaly is also prevented because of the blockchain consensus protocol. 
\section{Experiment}
\label{section: Experiment}

\subsection{Implementations}
\label{subsection: exp_implementation}
\textbf{SpendableStore.} SpendableStore's implementation consists of the bitcoin scripts for {\SpendableDO} and the off-chain {\SSInstance} server program.
We implemented \revise{the locking and unlocking scripts for KVP {\SpendableDO} (Section~\ref{section: SpendableDO}) at the opcode level~\cite{BSV2019opcodes} and applied a size optimization~\cite{Barbacovi2024pushtxbitshift}.} We implemented a prototype {\SSInstance} consisting of two programs corresponding to the Synchronizer and Transaction Manager (Section~\ref{subsection: Instance_Design}). 
The Synchronizer establishes persistent connections with the blockchain using JungleBus~\cite{JungleBus}.

The Transaction Manager program executes incoming {\SSTx}s serially. We achieve concurrent {\SSTx} processing by running multiple Transaction Manager programs and using Redis as the shared memory for all the programs.

\textbf{Baseline Implementations.}
\revise{
We established four baseline comparisons. (1) We compared our system with BlockchainDB~\cite{Hindi2019BlockchainDB} by implementing the Smart Contracts they designed and deploying them to the public Ethereum blockchain. (2) We also implemented BlockchainDB++, an augmented version, with two additional features to emulate {\SpendableStore}. First, we changed the BlockchainDB SC's "write-blockchain" method to take an array of key-value pairs instead of a single pair to support transactional multi-updates. Secondly, we included an owner key field for each key-value pair, similar to the ERC-20 protocol~\cite{ERC-20}, to provide data ownership protection. (3) We further modified BlockchainDB++ to operate on an Ethereum Layer-2 network. (4) We implemented the KVP {\SpendableDO} using Aiken~\cite{aiken2025cardano} to be compatible with the Cardano (ADA) Blockchain~\cite{cardano2025}, which utilizes an extended UTXO model similar to ours. Our selection of baselines aims to provide insights into how our system compares with those with an architecturally different blockchain layer (in cases 1, 2, and 3) and those with a similar architecture but limited off-chain capabilities (in case 4).
}

\subsection{Experiment Setup}
\label{subsection: exp_setup}
We conducted our experiments on AWS EC2 instances with 16 vCPUs and 32 GB of memory, running Ubuntu 24.04 LTS. For the Ethereum blockchain, we used the Holesky Testnet~\cite{Holesky_Testnet}. \revise{For Ethereum Layer-2, we used the Arbitrum Sepolia testnet~\cite{arbitrum2025sepolia}. For the Cardano blockchain, we used Cardano Preprod testnet~\cite{cardano2025testnet}. For {\SpendableStore}, we used the Bitcoin Satoshi Vision (BSV) Mainnet~\cite{BitcoinSV}. We did not choose the Ethereum Mainnet or the Bitcoin BTC Mainnet because of budget limitations. Although using a testnet gives different results than the Mainnet, they provide a sufficient framework for testing research prototypes and have been used by prior research studies~\cite{hu2018hierarchical, delgado2019txprobe, Hindi2019BlockchainDB, singh2023wedgeblock, ge2022hybrid}. \revise{Another reason for choosing BSV over BTC is that BSV does not impose a fixed limit on the block size, which allows us to experiment with large data objects.} \revise{We omitted testing against permissioned blockchain systems and less decentralized options to focus on the viability of a truly decentralized datastore.} Our experiments were recorded in the blockchain for verification\footnote{whatsonchain.com/address/1PwsXUKC8U1Ufmq8dNpM7ahFFX2WjjxJyv \\
 1GYmrMiVWLcKvNKe8uP6tU6Dc3JLJR3kGN \\
 1A3jX33UepqvJiZN79PHkGznDicJ7Rh6Vs \\
 17aW1JXXKoBZvarX67JXMkr4xHkAexnKmr}.

In Section~\ref{subsection: exp_vs_blockchainDB}, we \revise{showcase the experiment results and discuss the implications.} The workload used in this set of experiments is non-transactional, following the YCSB Benchmark~\cite{Cooper2010YCSB} standards because BlockchainDB does not support transactions. The measurement metrics we collected include the overall operation throughput, the latency distribution, and the monetary cost of each operation. 

In Sections~\ref{subsection: exp_heavy_Read} and~\ref{subsection: exp_heavy_write}, we evaluate {\SpendableStore}'s transaction processing with a read-heavy transaction workload and a write-heavy transaction workload, respectively. For each workload, we compare how {\SpendableStore} behaves under different FNSI levels. The measurement metrics are the overall transaction throughput and the transaction abort rate.

\begin{figure}[ht]
    \centering
    \includegraphics[width=\linewidth]{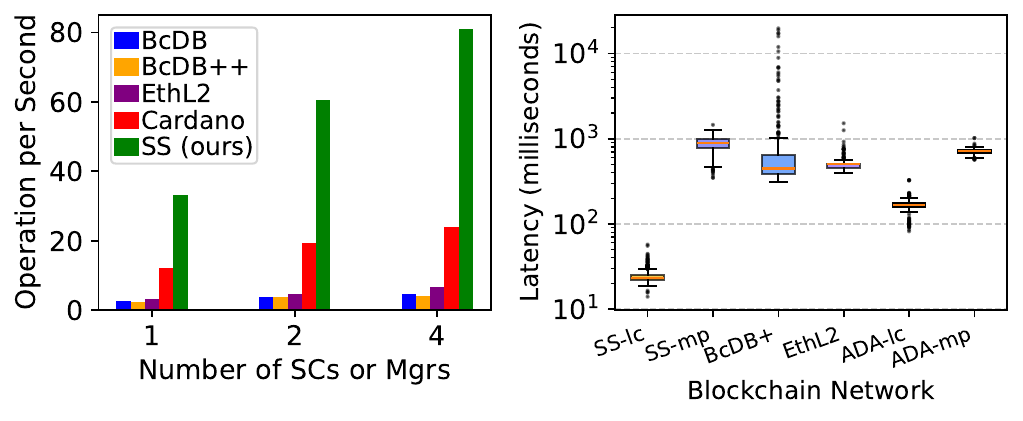}
    \caption{\revise{Baseline Comparison.}}
    \vspace{-1.5em}
    \label{fig: exp-G0}
\end{figure} 

\subsection{Baseline Comparison}
\label{subsection: exp_vs_blockchainDB}
\textbf{Experiment 1 (Fig.~\ref{fig: exp-G0}).} In this experiment, we compare the operation processing capability of \revise{5 blockchain-based database approaches---BlockchainDB (BcDB), BlockchainDB++ (BcDB++), BlockchainDB++ on Ethereum Layer-2 (EthL2), Cardano-based {\SpendableDO} (ADA), and {\SpendableStore} (SS) --- with a workload containing 1,000 conflict-free update operations.} We used this workload in favor of the Ethereum-based designs because the read operation in {\SpendableStore} is much faster as we leverage local caching of the blockchain, whereas they need to send read requests through the internet. 

We varied the parallelization level and recorded the total time it took for each design to complete the entire workload at the given level. We consider completing the workload as receiving all validation responses without waiting for the next block, since the block construction time is affected by factors irrelevant to our system designs. For ETH-based designs, a parallel level two means the workload is sharded into two SCs, and there are two corresponding processes sending write requests in parallel. For Cardano and {\SpendableStore}, a parallel level two means the workload is divided among two processes or transaction managers running in parallel. 

The left sub-figure of Fig.~\ref{fig: exp-G0} shows that {\SpendableStore} has around 16 times higher throughput than the Ethereum-based designs \revise{and 4 times higher throughput than the Cardano-based design.} The throughput for BcDB and its variations shows insignificant improvement with more SCs and remains below 5, which is consistent with the BlockchainDB paper's experiment result for the "native blockchain" case. BCDB++ has lower throughput than BcDB because its SC code logic is more complex and takes longer to execute. The low throughput for Ethereum-based designs is mainly due to the current Ethereum Virtual Machine executing all transactions serially~\cite{EthereumSerialExecution}. \revise{This applies to the Layer-2 network as well and explains the limited improvement.} During our experiments, we observed that if two requests to invoke the same SC method are sent without enough delay in between, the blockchain node treats the latter one as an attempt to replace the former and drops the former. This puts a major limitation on the throughput. On the other hand, the {\UTXO} design used by Cardano and {\SpendableStore} has better parallelism in validating incoming transactions~\cite{jiang2020bzip, muller2023reality} and does not have the request replacement issue as in Ethereum. Therefore, {\SpendableStore} transaction managers can send multiple transactions back-to-back, significantly improving the throughput. 

The right half of Fig.~\ref{fig: exp-G0} shows the latency distribution of write operations at parallel level 4. For BcDB++ and EthL2, the processing time is the time between sending the SC method call and receiving the corresponding transaction ID. \revise{For Cardano and {\SpendableStore}, we took two measurements after the start of the operation: one for constructing the {\UTXO} transaction locally (-lc) and another for receiving mempool validation (-mp).} Since a local {\UTXO} transaction has a deterministic outcome, it can be used as an early response to improve user experience. 

We observed that with BcDB++, the first 90\% of the write operations are processed within 1 second; however, the remaining 10\% took as long as over 10 seconds. This suggests that the layer-1 nodes are overloaded near the end. \revise{EthL2 has a lighter tail because the Layer-2 nodes experience a more moderate overall workload.} In comparison, the {\SpendableStore} receives most of the operations' mempool validations with longer waiting; however, the worst response time did not exceed 1.5 seconds. Moreover, {\SpendableStore} can provide early responses in the magnitude of single to double-digit milliseconds. These show that {\SpendableStore} can provide better user experiences than BcDB++ under heavy workloads. \revise{Although the Cardano-based approach also enjoys early local responses, they are not comparable to {\SpendableStore} because the extended UTXO model restrains one from accessing the low-level scripts in exchange for better programmability, thus preventing significant off-chain node optimizations.} 

\begin{table}[t]
\centering
\setlength{\tabcolsep}{3pt}
\caption{\revise{Cost Analysis.}}
\label{tab:blockchain_metrics}
\begin{tabular}{|c|c|c||c|c||c|c|c|}
\hline
\multirow{2}{*}{\textbf{}} & \multicolumn{2}{c||}{\textbf{Deploy}} & \multicolumn{2}{c||}{\textbf{Create}} & \multicolumn{2}{c|}{\textbf{Update}} & \multirow{2}{*}{\textbf{Per 1KB}} \\ \cline{2-7}
        &Bytes & USD   &Bytes& USD    & Bytes& USD      &        \\ \hline \hline
SS      & N/A  & N/A   & 508 & 1E-05  & 1120 & 2.3E-05  &  2E-05 \\ \hline
BcDB    & 2669 & 0.089 & 196 & 0.162  & N/A  & N/A      &  2.22  \\ \hline
BcDB++  & 5614 & 0.183 & 228 & 0.33   & 324  & 0.21     &  2.22  \\ \hline
EthL2   & 5614 & 0.079 & 228 & 4.8E-03& 324  & 3.3E-03  &  0.042 \\ \hline
Cardano & N/A  & N/A   & 400 & 0.085  & 1270 & 0.415    &  2.15  \\ \hline
Ordinal & N/A  & N/A   & 151 & 3E-6   & N/A  & N/A      &  2E-05 \\ \hline
\end{tabular}
\vspace{-1em}
\end{table}

\revise{We analyze the costs of different operations for each baseline in Table~\ref{tab:blockchain_metrics}. The deploy column refers to the allocation of a Smart Contract Account and is applicable only to the Ethereum-based methods. The create column is the overhead for performing a data insertion. The update column is the overhead for performing an atomic update on a set of inserted data. BcDB does not support atomic updates. Each column has two sub-columns, one for the size of the instruction script that will be processed by the blockchain to perform the operation, the other for the monetary cost in USD. The Per-1KB column is the additional cost for each 1KB of data to be inserted or updated to. We also included Ordinals~\cite{Ordinals} as it provides the minimum overhead for recording data on a UTXO blockchain, without any functionalities like transactional updates and access control. We calculated the dollar amounts using \$3,000, \$0.5, and \$20 per coin for Ethereum, Cardano, and BSV, respectively, based on their historical averages. 
}

\revise{
Our approach has a lower cost on most metrics than the other baselines. When compared with the Ordinals, we achieve transactional abilities and access controls efficiently with only 357 additional bytes in the script design. If calculated using the BTC's price at \$100,000 dollars per coin instead, our costs would be 0.051, 0.11, and 0.1 respectively. They still remain at a comparable level to the others. 
}

\begin{figure}[ht]
    \centering
    \includegraphics[width=0.8\linewidth]{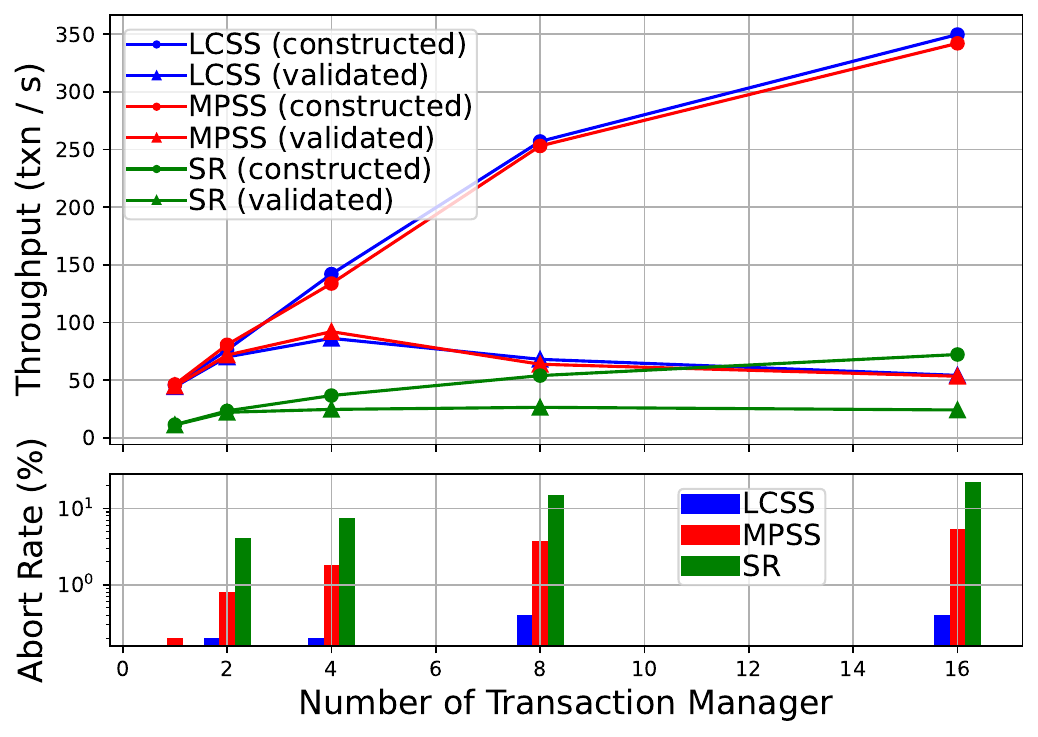}
    \caption{Read Heavy Workload (var. Concurrency Level).}
    \vspace{-1.5em}
    \label{fig: exp-G3}
\end{figure} 

\begin{figure}[ht]
    \centering
    \includegraphics[width=0.8\linewidth]{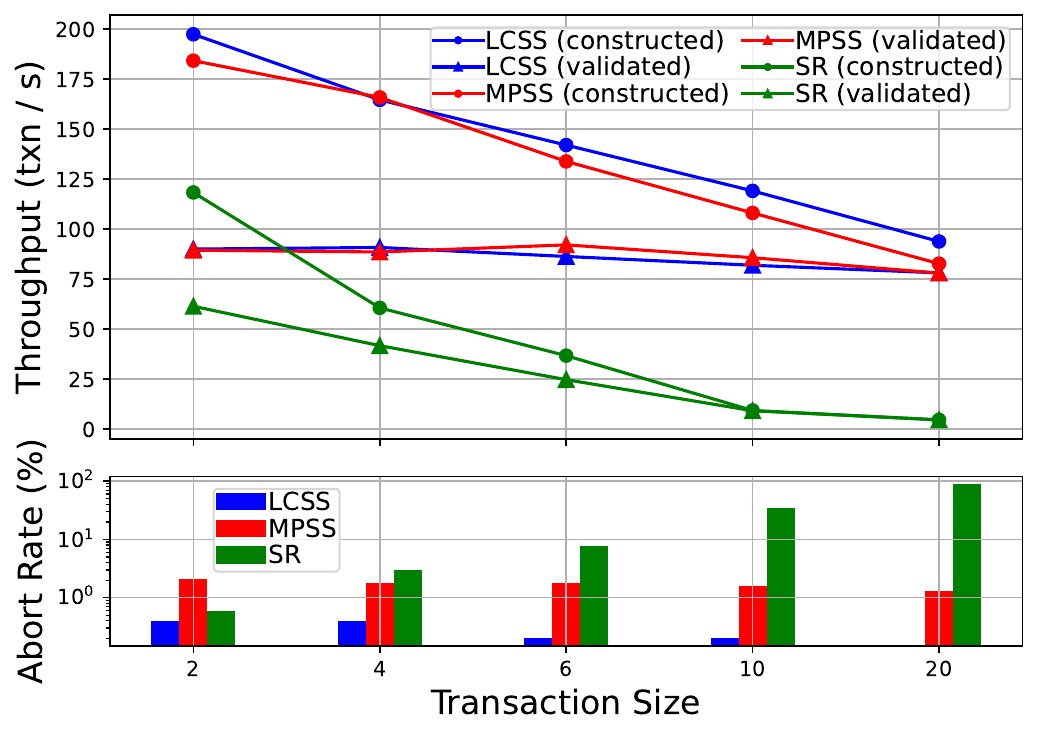}
    \caption{Read Heavy Workload (var. Txn Size).}
    \vspace{-1.3em}
    \label{fig: exp-G4}
\end{figure} 

\subsection{Read-Heavy Workload}
\label{subsection: exp_heavy_Read}
In the following two experiments, we evaluate {\SpendableStore}'s transaction processing ability with read-heavy workloads that contain a small amount of write-write conflicting {\SSTx}s. To prepare for the workloads, we populated the database by deploying 5000 {\SpendableDO}s to the blockchain. We define \textit{X} as the size of a {\SSTx}. We make each {\SSTx} in the workload read \textit{X-1} {\SpendableDO}s and write \textit{1} {\SpendableDO} for a total of \textit{X} accessed {\SpendableDO}s, all chosen uniformly at random from the 5000. Each workload consists of 512 different {\SSTx} of size \textit{X}. 

\textbf{Experiment 2a (Fig.~\ref{fig: exp-G3}).} In this experiment, we fixed the {\SSTx} size at 6, which translates to 83\% read and 17\% write in the workload, and evaluated the {\SpendableStore}'s scalability by doubling the number of transaction managers working in parallel from 1 to 16. With each setting, we run the workload three times, setting all transaction isolation levels to Serializable (SR), Mempool Snapshot Isolation (MPSS), and Local Snapshot Isolation (LCSS), respectively. We omitted to test Blockchain Snapshot Isolation (BCSS) because it is only suitable for read-only {\SSTx}. In each experiment run, we measure the time to finish constructing all transactions' BC-Twin locally and the time to receive all their validation responses.  We calculate two throughput numbers -- "construction throughput" and "validation throughput" -- based on the two measurements. We also record an abort rate as the percentage of {\SSTx}s that were not validated by the mining nodes. 

Fig.~\ref{fig: exp-G3} shows that LCSS and MPSS have a similar performance, which is much better than SR. Under LCSS and MPSS, the constructed BC-Twin for each {\SSTx} has only one input-output {\oVersion} pair, whereas under SR, the BC-Twin has 6 pairs. The higher complexity in the BC-Twin prolongs both the construction and the validation time. It also spikes up the abort rate since any read-write conflicts will be invalidated by the blockchain validation nodes. 

The figure also shows that only the "construction throughput" scales with more transaction managers, while the "validation throughput" is marginally and even negatively impacted. With four transaction managers, the validation throughput peaked at around 90 TPS. With more transaction managers, although we can send constructed BC-Twins to the blockchain at a higher rate, the validation nodes seem to be overloaded and take a longer time to finish all validations. This indicates that the bottleneck in our current experiment setup is in the blockchain validation nodes. 

\textbf{Experiment 2b (Fig.~\ref{fig: exp-G4}).} In this experiment, we fixed the number of transaction managers at 4 and compared SR, LCSS, and MPSS at {\SSTx} size  2, 4, 6, 10, and 20.

As {\SSTx} size increases, the abort rate quickly increases for SR for the same reason we explained in the previous experiment. However, the abort rate for LCSS and MPSS slightly decreases. The reason is that since each transaction manager took a longer time to process each {\SSTx}, the probability of an {\SSTx} observing a stale snapshot, which can cause it to be aborted, becomes lower.

The construction throughput continues to decrease for all isolation levels as expected. The validation throughput significantly decreases for SR because the blockchain validation node has to spend a longer time validating each BC-Twin. For LCSS and MPSS, the validation throughput is not affected much because the BC-Twin size remains at 1. 

Experiments 2a and 2b show that the two FNSI isolation levels outperform serializable execution for read-heavy workloads at a cheaper cost. For an {\SSTx} of size 20, SR would be 20 times more expensive. Furthermore, since performing reads using an LCSS is prone to dirty reads and non-repeatable read anomalies, we conclude that MPSS is the best snapshot isolation level for read-heavy workloads.

\begin{figure}[t]
    \centering
    \includegraphics[width=0.8\linewidth]{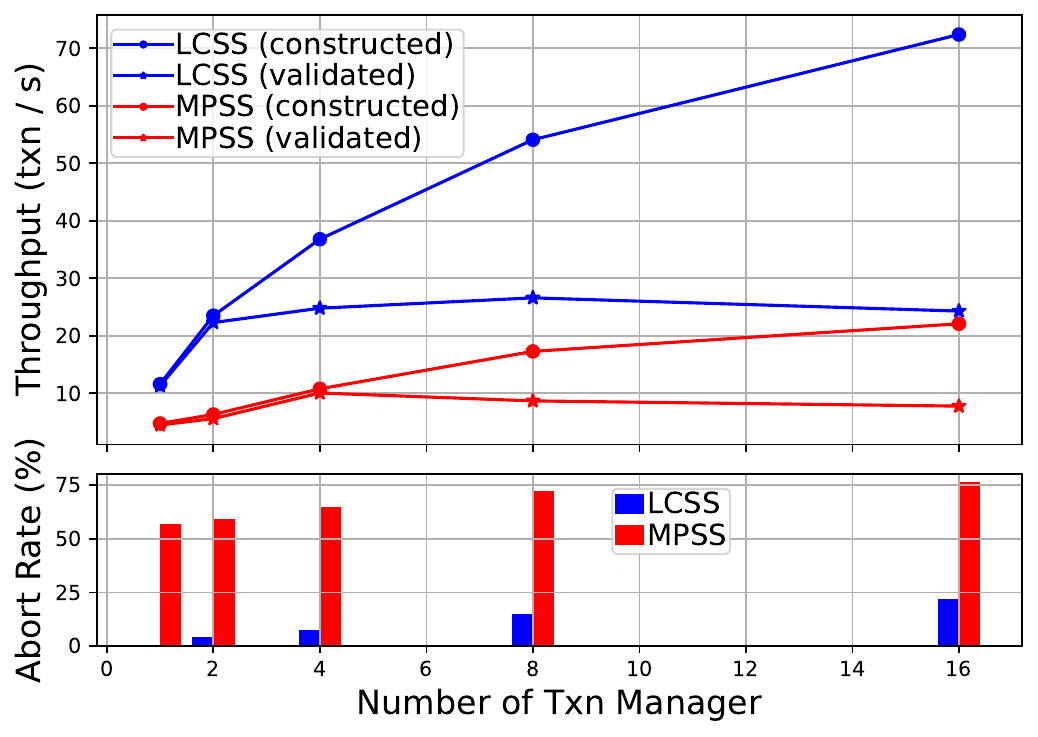}
    \caption{Write Heavy Workload (var. Concurrency Level).}
    \vspace{-1.7em}
    \label{fig: exp-G1}
\end{figure} 

\begin{figure}[ht]
    \centering
    \includegraphics[width=0.8\linewidth]{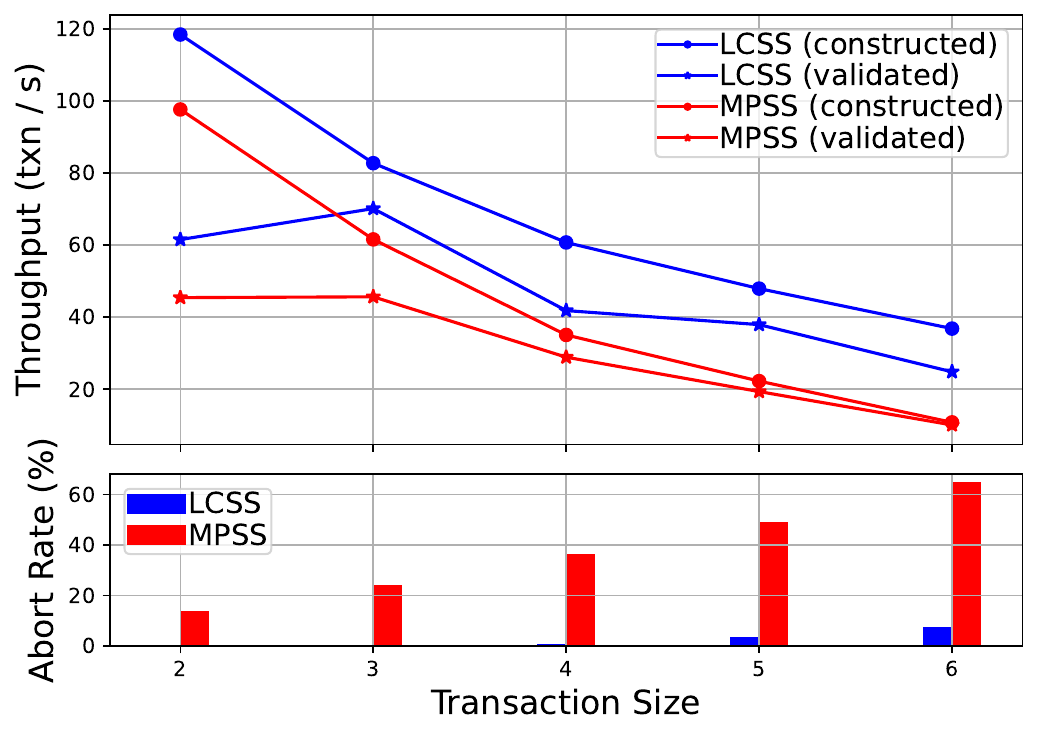}
    \caption{Write Heavy Workload (var. Txn Size).}
    \vspace{-1.7em}
    \label{fig: exp-G2}
\end{figure} 

\subsection{Write-Heavy Workload}
\label{subsection: exp_heavy_write}
In the following two experiments, we evaluate {\SpendableStore}'s transaction processing ability with write-heavy workloads that contain many write-write conflicting {\SSTx}s. The database has the same 5000 {\SpendableDO}s as in the previous experiments. We make each {\SSTx} in the workload perform a read followed by a write on \textit{X} number of {\SpendableDO}s chosen uniformly at random from the 5000, i.e., its BC-Twin has \textit{X} input-output {\oVersion} pairs. Each workload consists of 512 different {\SSTx} of size \textit{X}. In each experiment run, we measure "construction throughput", "validation throughput", and total abort rate the same way as in the read-heavy workloads.

\textbf{Experiment 3a (Fig.~\ref{fig: exp-G1}).} In this experiment, we fixed the {\SSTx} size at 6 and evaluated the {\SpendableStore}'s scalability by doubling the number of transaction managers working in parallel from 1 to 16. With each setting, we run the workload twice, setting all transactions' isolation levels to MPSS and LCSS, respectively. There is no Serializable option because these {\SSTx} are already Serializable by \textit{spending} all their reads. 

Fig.~\ref{fig: exp-G1} shows that LCSS performs better than MPSS for write-heavy workloads. It has a lower abort rate and, as a result, higher overall throughput. With one transaction manager using LCSS, there is a 0\% abort rate. For throughput, we observe that only the "construction throughput" scales with more transaction managers, while the "validation throughput" is marginally and even negatively impacted. This shows that the bottleneck in our current experiment setup is in the blockchain validation nodes.

\textbf{Experiment 3b (Fig.~\ref{fig: exp-G2}).} In this experiment, we fixed the number of transaction managers at 4 and varied the {\SSTx} size from 2 to 6. As the size of {\SSTx} increases, more write-write conflicts are present in the workload; therefore, the overall abort rate increases while the throughput decreases. However, there is an exception at size 2 where the "validation throughput" is not higher than that of size 3. This shows that at size 2, our transaction managers overloaded the blockchain validation node. Experiments 3a and 3b show that LCSS outperforms MPSS for write-heavy workloads.
\vspace{-0.1em}
\section{Related Works}
\label{section: Related Works}

\textbf{Blockchain-based Databases.}
There has been prior work on building blockchain-based databases~\cite{Hindi2019BlockchainDB,mcconaghy2016bigchaindb,TendermintDiscontinue,CovenantSQL,yan2021blockchain, nathan2019blockchain, peng2020falcondb, gupta2020resilientdb, allen2019veritas, Lin2023RollStore}.
BigchainDB~\cite{mcconaghy2016bigchaindb} (Version 2.0) is built on top of Tendermint~\cite{buchman2016tendermint}, and it adds a blockchain layer to a traditional database design. CovenantSQL~\cite{CovenantSQL} aims to establish a blockchain architecture on top of a SQL database. It runs on a self-implemented global consensus layer instead of using existing blockchain networks. Other similar approaches with permissioned blockchain settings include~\cite{yan2021blockchain, nathan2019blockchain, peng2020falcondb, gupta2020resilientdb, allen2019veritas, Gu2023MLonChain}, they all offer less data decentralization than {\SpendableStore}.



\textbf{Trusted Execution Environment.}
Some works~\cite{mast2019vision, xie2023tebds, wang2020spds} explore adding blockchain properties to data sharing using a trusted execution environment. They focus on achieving high verifiability while avoiding the computational overheads of using the blockchain. They do not focus on providing better decentralization in terms of data ownership. They are also limited to simple operations without transactional support.

\textbf{NFT.}
Non-fungible tokens were made popular by Ethereum with the ERC-721 standard~\cite{ERC-721}. They are good examples of decentralized applications that support data ownership over digital assets. However, they are designed to hold static data, such as a hash value of an image or a URL that redirects to the IPFS~\cite{IPFS} network hosting the image. They are not designed to support the update operation and cannot be directly used for database designs. Dynamic NFT (dNFT) under the ERC-1155 standard~\cite{ERC-1155} is an improvement that allows for updates. However, data contained in a dNFT is still bound to a specific Ethereum account, which restricts the generality of the multi-update transactions. Whereas in {\SpendableStore}, the flexibility of {\UTXO}-transactions ensures a one-to-one matching from a database transaction ({\SSTx}) to a Bitcoin transaction (BC-Twin) (see Section~\ref{subsection: SDO_multi_updates}).

\textbf{Ordinal.}
Also known as a {\UTXO} inscription or a {\UTXO} NFT, Ordinal~\cite{Ordinals} is a recent method to record arbitrary data on the {\UTXO} blockchain. Ordinals are transaction outputs that put data after a permanently locked locking script. Ordinal outputs are provably unspendable, thus not updateable, because their locking scripts start with a specific set of opcodes that always make the validation algorithm return false no matter what unlocking script is provided. {\SpendableStore} adds the update functionality to Ordinals by making them spendable.

\vspace{-0.1em}
\section{Conclusion}
\label{section: Conclusion}

In this paper, we presented {\SpendableStore}, the first blockchain-based database that builds on a {\UTXO}-based blockchain. The aim of {\SpendableStore} is to achieve data decentralization and data ownership. This is achieved by a design that allows data owners to control their data objects fully without relying on a centralized entity. {\SpendableStore} supports database transaction processing and proposes a new isolation guarantee called Future Now Snapshot Isolation (FNSI) that allows programmers to balance between the correctness of concurrent database transactions on {\SpendableStore} while maintaining the performance benefits of using a {\UTXO}-based blockchain. Our experiments on a public blockchain Mainnet demonstrate that {\SpendableStore} can achieve up to 16x better throughput compared to a state-of-the-art blockchain-based database while providing database transactions and decentralized ownership protection.

\section{AI-Generated Content Acknowledgment}
\label{section: AI_Acknowledgement}

The authors confirm that no artificial intelligence (AI)–based tools were used to generate or assist in the writing of this manuscript. All content was produced by the authors.


\bibliographystyle{unsrturl}
\bibliography{citations}

\begin{thebibliography}{10}

\bibitem{nakamoto2008bitcoin}
Satoshi Nakamoto.
\newblock Bitcoin: A peer-to-peer electronic cash system.
\newblock {\em Decentralized business review}, 2008.

\bibitem{lamport2019byzantine}
Leslie Lamport, Robert Shostak, and Marshall Pease.
\newblock The byzantine generals problem.
\newblock In {\em Concurrency: the works of leslie lamport}, pages 203--226. 2019.

\bibitem{korth2021notes}
Henry~F Korth.
\newblock Notes on blockchain database systems.
\newblock 2021.

\bibitem{zamani2018rapidchain}
Mahdi Zamani, Mahnush Movahedi, and Mariana Raykova.
\newblock Rapidchain: Scaling blockchain via full sharding.
\newblock In {\em Proceedings of the 2018 ACM SIGSAC conference on computer and communications security}, pages 931--948, 2018.

\bibitem{Fang2022PeloPartition}
Juncheng Fang, Farzad Habibi, Kevin Bruhwiler, Fayzah Alshammari, Abhishek Singh, Yinan Zhou, and Faisal Nawab.
\newblock Pelopartition: Improving blockchain resilience to network partitioning.
\newblock In {\em 2022 IEEE International Conference on Blockchain (Blockchain)}, pages 274--281, 2022.
\newblock \href {https://doi.org/10.1109/Blockchain55522.2022.00045} {\path{doi:10.1109/Blockchain55522.2022.00045}}.

\bibitem{singh2020sidechain}
Amritraj Singh, Kelly Click, Reza~M Parizi, Qi~Zhang, Ali Dehghantanha, and Kim-Kwang~Raymond Choo.
\newblock Sidechain technologies in blockchain networks: An examination and state-of-the-art review.
\newblock {\em Journal of Network and Computer Applications}, 149:102471, 2020.

\bibitem{xu2021slimchain}
Cheng Xu, Ce~Zhang, Jianliang Xu, and Jian Pei.
\newblock Slimchain: Scaling blockchain transactions through off-chain storage and parallel processing.
\newblock {\em Proceedings of the VLDB Endowment}, 14(11):2314--2326, 2021.

\bibitem{tan2020latte}
Sean Tan, Sourav S~Bhowmick, Huey~Eng Chua, and Xiaokui Xiao.
\newblock Latte: Visual construction of smart contracts.
\newblock In {\em Proceedings of the 2020 ACM SIGMOD International Conference on Management of Data}, pages 2713--2716, 2020.

\bibitem{bragagnolo2018smartinspect}
Santiago Bragagnolo, Henrique Rocha, Marcus Denker, and St{\'e}phane Ducasse.
\newblock Smartinspect: solidity smart contract inspector.
\newblock In {\em 2018 International workshop on blockchain oriented software engineering (IWBOSE)}, pages 9--18. Ieee, 2018.

\bibitem{zahnentferner2018chimeric}
Joachim Zahnentferner.
\newblock Chimeric ledgers: Translating and unifying utxo-based and account-based cryptocurrencies.
\newblock {\em Cryptology ePrint Archive}, 2018.

\bibitem{brunjes2020utxo}
Lars Br{\"u}njes and Murdoch~J Gabbay.
\newblock Utxo-vs account-based smart contract blockchain programming paradigms.
\newblock In {\em Leveraging Applications of Formal Methods, Verification and Validation: Applications: 9th International Symposium on Leveraging Applications of Formal Methods, ISoLA 2020, Rhodes, Greece, October 20--30, 2020, Proceedings, Part III 9}, pages 73--88. Springer, 2020.

\bibitem{baur2021volatility}
Dirk~G Baur and Thomas Dimpfl.
\newblock The volatility of bitcoin and its role as a medium of exchange and a store of value.
\newblock {\em Empirical Economics}, 61(5):2663--2683, 2021.

\bibitem{kubat2015virtual}
Max Kub{\'a}t.
\newblock Virtual currency bitcoin in the scope of money definition and store of value.
\newblock {\em Procedia Economics and Finance}, 30:409--416, 2015.

\bibitem{mattke2020cryptocurrency}
Jens Mattke, Christian Maier, and Lea Reis.
\newblock Is cryptocurrency money? three empirical studies analyzing medium of exchange, store of value and unit of account.
\newblock In {\em Proceedings of the 2020 on Computers and People Research Conference}, pages 26--35, 2020.

\bibitem{gurtu2019potential}
Amulya Gurtu and Jestin Johny.
\newblock Potential of blockchain technology in supply chain management: a literature review.
\newblock {\em International Journal of Physical Distribution \& Logistics Management}, 49(9):881--900, 2019.

\bibitem{saberi2019blockchain}
Sara Saberi, Mahtab Kouhizadeh, Joseph Sarkis, and Lejia Shen.
\newblock Blockchain technology and its relationships to sustainable supply chain management.
\newblock {\em International journal of production research}, 57(7):2117--2135, 2019.

\bibitem{truong2023blockchain}
Vu~Tuan Truong, Long Le, and Dusit Niyato.
\newblock Blockchain meets metaverse and digital asset management: A comprehensive survey.
\newblock {\em Ieee Access}, 11:26258--26288, 2023.

\bibitem{zakhary2019towards}
Victor Zakhary, Mohammad~Javad Amiri, Sujaya Maiyya, Divyakant Agrawal, and Amr~El Abbadi.
\newblock Towards global asset management in blockchain systems.
\newblock {\em arXiv preprint arXiv:1905.09359}, 2019.

\bibitem{baltais2024economic}
Markuss Baltais, Evita Sondore, Talis~J Putni{\c{n}}{\v{s}}a, and Jonathan~R Karlsen.
\newblock Economic impact potential of real-world asset tokenization.
\newblock {\em UTS Business School, University of Technology Sydney, Report}, pages 2024--06, 2024.

\bibitem{reyna2018blockchain}
Ana Reyna, Cristian Mart{\'\i}n, Jaime Chen, Enrique Soler, and Manuel D{\'\i}az.
\newblock On blockchain and its integration with iot. challenges and opportunities.
\newblock {\em Future generation computer systems}, 88:173--190, 2018.

\bibitem{Maha2019BCIOT}
Maha Alaslani, Faisal Nawab, and Basem Shihada.
\newblock Blockchain in iot systems: End-to-end delay evaluation.
\newblock {\em IEEE Internet of Things Journal}, 6(5):8332--8344, 2019.
\newblock \href {https://doi.org/10.1109/JIOT.2019.2917226} {\path{doi:10.1109/JIOT.2019.2917226}}.

\bibitem{agbo2019blockchain}
Cornelius~C Agbo, Qusay~H Mahmoud, and J~Mikael Eklund.
\newblock Blockchain technology in healthcare: a systematic review.
\newblock In {\em Healthcare}, volume~7, page~56. MDPI, 2019.

\bibitem{zhang2018blockchain}
Peng Zhang, Douglas~C Schmidt, Jules White, and Gunther Lenz.
\newblock Blockchain technology use cases in healthcare.
\newblock In {\em Advances in computers}, volume 111, pages 1--41. Elsevier, 2018.

\bibitem{hadfield2025economy}
Gillian~K Hadfield and Andrew Koh.
\newblock An economy of ai agents.
\newblock {\em arXiv preprint arXiv:2509.01063}, 2025.

\bibitem{zhang2021pushtx}
Zhang.
\newblock Pushtx and its building blocks.
\newblock https://nchain.com/wp-content/uploads/2022/03/WP1605{\_}PUSHTX-and-its-Building-Blocks.pdf, 2021.
\newblock Accessed: 2026-01-20.

\bibitem{Hindi2019BlockchainDB}
Muhammad El-Hindi, Carsten Binnig, Arvind Arasu, Donald Kossmann, and Ravi Ramamurthy.
\newblock Blockchaindb: a shared database on blockchains.
\newblock {\em Proc. VLDB Endow.}, 12(11):1597–1609, July 2019.
\newblock \href {https://doi.org/10.14778/3342263.3342636} {\path{doi:10.14778/3342263.3342636}}.

\bibitem{PushdataOpcodes}
BitcoinSV Community.
\newblock Pushdata opcodes, 2022.
\newblock Accessed: 2024-10-15.
\newblock URL: \url{https://wiki.bitcoinsv.io/index.php/Pushdata_Opcodes}.

\bibitem{eberhardt2017or}
Jacob Eberhardt and Stefan Tai.
\newblock On or off the blockchain? insights on off-chaining computation and data.
\newblock In {\em Service-Oriented and Cloud Computing: 6th IFIP WG 2.14 European Conference, ESOCC 2017, Oslo, Norway, September 27-29, 2017, Proceedings 6}, pages 3--15. Springer, 2017.

\bibitem{Sovran2011PSI}
Yair Sovran, Russell Power, Marcos~K. Aguilera, and Jinyang Li.
\newblock Transactional storage for geo-replicated systems.
\newblock In {\em Proceedings of the Twenty-Third ACM Symposium on Operating Systems Principles}, SOSP '11, page 385–400, New York, NY, USA, 2011. Association for Computing Machinery.
\newblock \href {https://doi.org/10.1145/2043556.2043592} {\path{doi:10.1145/2043556.2043592}}.

\bibitem{elnikety2004generalized}
Sameh Elnikety, Fernando Pedone, and Willy Zwaenepoel.
\newblock Generalized snapshot isolation and a prefix-consistent implementation.
\newblock 2004.

\bibitem{BSV2019opcodes}
BSV Community.
\newblock Opcodes used in bitcoin script, 2019.
\newblock Accessed: 2026-01-20.
\newblock URL: \url{https://wiki.bitcoinsv.io/index.php/Opcodes\_used\_in\_Bitcoin\_Script}.

\bibitem{Barbacovi2024pushtxbitshift}
Barbacovi.
\newblock Op\_checksig beyong signature validation: a more efficient pushtx, 2024.
\newblock Accessed: 2026-01-20.
\newblock URL: \url{https://hackmd.io/@federicobarbacovi/H1DqEzfm1l}.

\bibitem{JungleBus}
Gorilla Pool.
\newblock Gorillapool junglebus, 2024.
\newblock Accessed: 2026-01-20.
\newblock URL: \url{https://junglebus.gorillapool.io/}.

\bibitem{ERC-20}
Ethereum Community.
\newblock Erc-20 token standard, 2024.
\newblock Accessed: 2026-01-20.
\newblock URL: \url{https://ethereum.org/en/developers/docs/standards/tokens/erc-20/}.

\bibitem{aiken2025cardano}
Cardano Community.
\newblock The modern smart contract platform for cardano, 2020.
\newblock Accessed: 2026-01-20.
\newblock URL: \url{https://aiken-lang.org/}.

\bibitem{cardano2025}
Cardano Community.
\newblock Cardano doc, 2020.
\newblock Accessed: 2026-01-20.
\newblock URL: \url{https://cardano.org/}.

\bibitem{Holesky_Testnet}
Ethereum Community.
\newblock Holesky testnet, 2024.
\newblock Accessed: 2024-10-15.
\newblock URL: \url{https://holesky.etherscan.io/}.

\bibitem{arbitrum2025sepolia}
Arbitrum Community.
\newblock Arbitrum chains overview, 2020.
\newblock Accessed: 2026-01-20.
\newblock URL: \url{https://docs.arbitrum.io/build-decentralized-apps/public-chains}.

\bibitem{cardano2025testnet}
Cardano Community.
\newblock Cardano testnets, 2022.
\newblock Accessed: 2026-01-20.
\newblock URL: \url{https://developers.cardano.org/docs/get-started/networks/testnets/}.

\bibitem{BitcoinSV}
BitcoinSV Community.
\newblock Bsv blockchain, 2024.
\newblock Accessed: 2024-10-15.
\newblock URL: \url{https://www.bsvblockchain.org/}.

\bibitem{hu2018hierarchical}
Yao-Chieh Hu, Ting-Ting Lee, Dimitris Chatzopoulos, and Pan Hui.
\newblock Hierarchical interactions between ethereum smart contracts across testnets.
\newblock In {\em Proceedings of the 1st Workshop on Cryptocurrencies and Blockchains for Distributed Systems}, pages 7--12, 2018.

\bibitem{delgado2019txprobe}
Sergi Delgado-Segura, Surya Bakshi, Cristina P{\'e}rez-Sol{\`a}, James Litton, Andrew Pachulski, Andrew Miller, and Bobby Bhattacharjee.
\newblock Txprobe: Discovering bitcoin’s network topology using orphan transactions.
\newblock In {\em Financial Cryptography and Data Security: 23rd International Conference, FC 2019, Frigate Bay, St. Kitts and Nevis, February 18--22, 2019, Revised Selected Papers 23}, pages 550--566. Springer, 2019.

\bibitem{singh2023wedgeblock}
Abhishek~A Singh, Yinan Zhou, Mohammad Sadoghi, Sharad Mehrotra, Shantanu Sharma, and Faisal Nawab.
\newblock Wedgeblock: An off-chain secure logging platform for blockchain applications.
\newblock In {\em EDBT}, pages 526--539, 2023.

\bibitem{ge2022hybrid}
Zerui Ge, Dumitrel Loghin, Beng~Chin Ooi, Pingcheng Ruan, and Tianwen Wang.
\newblock Hybrid blockchain database systems: design and performance.
\newblock {\em Proc. VLDB Endow.}, 15(5):1092–1104, January 2022.
\newblock \href {https://doi.org/10.14778/3510397.3510406} {\path{doi:10.14778/3510397.3510406}}.

\bibitem{Cooper2010YCSB}
Brian~F. Cooper, Adam Silberstein, Erwin Tam, Raghu Ramakrishnan, and Russell Sears.
\newblock Benchmarking cloud serving systems with ycsb.
\newblock In {\em Proceedings of the 1st ACM Symposium on Cloud Computing}, SoCC '10, page 143–154, New York, NY, USA, 2010. Association for Computing Machinery.
\newblock URL: \url{https://doi.org/10.1145/1807128.1807152}.

\bibitem{EthereumSerialExecution}
Fuel Labs.
\newblock Ethereum’s scalability crisis: The execution layer, 2024.
\newblock Accessed: 2024-10-15.
\newblock URL: \url{https://fuel.mirror.xyz/uQxyb1o_Gu4oBSyT1ULuqRu7ffmIXuZtx9ux8ndFMXs}.

\bibitem{jiang2020bzip}
Shuhao Jiang, Jiajun Li, Shijun Gong, Junchao Yan, Guihai Yan, Yi~Sun, and Xiaowei Li.
\newblock Bzip: A compact data memory system for utxo-based blockchains.
\newblock {\em Journal of Systems Architecture}, 109:101809, 2020.

\bibitem{muller2023reality}
Sebastian M{\"u}ller, Andreas Penzkofer, Nikita Polyanskii, Jonas Theis, William Sanders, and Hans Moog.
\newblock Reality-based utxo ledger.
\newblock {\em Distributed Ledger Technologies: Research and Practice}, 2(3):1--33, 2023.

\bibitem{Ordinals}
Kyle Torpey.
\newblock Bitcoin ordinal nft: Everything you need to know, 2024.
\newblock Accessed: 2024-10-15.
\newblock URL: \url{https://www.investopedia.com/what-are-bitcoin-ordinals-7486436}.

\bibitem{mcconaghy2016bigchaindb}
Trent McConaghy, Rodolphe Marques, Andreas M{\"u}ller, Dimitri De~Jonghe, Troy McConaghy, Greg McMullen, Ryan Henderson, Sylvain Bellemare, and Alberto Granzotto.
\newblock Bigchaindb: a scalable blockchain database.
\newblock {\em white paper, BigChainDB}, pages 53--72, 2016.

\bibitem{TendermintDiscontinue}
Interchain.
\newblock Discontinuing tendermint core v0.35: A postmortem on the new networking layer, 2022.
\newblock Accessed: 2024-10-15.
\newblock URL: \url{https://medium.com/the-interchain-foundation/discontinuing-tendermint-v0-35-a-postmortem-on-the-new-networking-layer-3696c811dabc}.

\bibitem{CovenantSQL}
CovenantSQL Community.
\newblock Covenantsql, 2019.
\newblock Accessed: 2024-10-15.
\newblock URL: \url{https://github.com/CovenantSQL/CovenantSQL}.

\bibitem{yan2021blockchain}
Dekai Yan, Xiaohua Jia, Jiangang Shu, and Rutao Yu.
\newblock A blockchain-based database system for decentralized information management.
\newblock In {\em 2021 IEEE Global Communications Conference (GLOBECOM)}, pages 1--6. IEEE, 2021.

\bibitem{nathan2019blockchain}
Senthil Nathan, Chander Govindarajan, Adarsh Saraf, Manish Sethi, and Praveen Jayachandran.
\newblock Blockchain meets database: Design and implementation of a blockchain relational database.
\newblock {\em arXiv preprint arXiv:1903.01919}, 2019.

\bibitem{peng2020falcondb}
Yanqing Peng, Min Du, Feifei Li, Raymond Cheng, and Dawn Song.
\newblock Falcondb: Blockchain-based collaborative database.
\newblock In {\em Proceedings of the 2020 ACM SIGMOD international conference on management of data}, pages 637--652, 2020.

\bibitem{gupta2020resilientdb}
Suyash Gupta, Sajjad Rahnama, Jelle Hellings, and Mohammad Sadoghi.
\newblock Resilientdb: Global scale resilient blockchain fabric.
\newblock {\em arXiv preprint arXiv:2002.00160}, 2020.

\bibitem{allen2019veritas}
Lindsey Allen, Panagiotis Antonopoulos, Arvind Arasu, Johannes Gehrke, Joachim Hammer, James Hunter, Raghav Kaushik, Donald Kossmann, Jonathan Lee, Ravi Ramamurthy, et~al.
\newblock Veritas: Shared verifiable databases and tables in the cloud.
\newblock In {\em 9th Biennial Conference on Innovative Data Systems Research (CIDR)}, pages 1--9, 2019.

\bibitem{Lin2023RollStore}
Qi~Lin, Binbin Gu, and Faisal Nawab.
\newblock Rollstore: Hybrid onchain-offchain data indexing for blockchain applications.
\newblock {\em IEEE Transactions on Knowledge and Data Engineering}, 36(12):9176--9191, 2024.
\newblock \href {https://doi.org/10.1109/TKDE.2024.3436514} {\path{doi:10.1109/TKDE.2024.3436514}}.

\bibitem{buchman2016tendermint}
Ethan Buchman.
\newblock {\em Tendermint: Byzantine fault tolerance in the age of blockchains}.
\newblock PhD thesis, University of Guelph, 2016.

\bibitem{Gu2023MLonChain}
Binbin Gu, Abhishek Singh, Yinan Zhou, Juncheng Fang, and Faisal Nawab.
\newblock Ml on chain: The case and taxonomy of machine learning on blockchain.
\newblock In {\em 2023 IEEE International Conference on Blockchain and Cryptocurrency (ICBC)}, pages 1--18, 2023.
\newblock \href {https://doi.org/10.1109/ICBC56567.2023.10174908} {\path{doi:10.1109/ICBC56567.2023.10174908}}.

\bibitem{mast2019vision}
Kai Mast, Lequn Chen, and Emin~G{\"u}n Sirer.
\newblock A vision for autonomous blockchains backed by secure hardware.
\newblock In {\em Proceedings of the 4th Workshop on System Software for Trusted Execution}, pages 1--6, 2019.

\bibitem{xie2023tebds}
Hui Xie, Jun Zheng, Teng He, Shengjun Wei, and Changzhen Hu.
\newblock Tebds: A trusted execution environment-and-blockchain-supported iot data sharing system.
\newblock {\em Future Generation Computer Systems}, 140:321--330, 2023.

\bibitem{wang2020spds}
Yuntao Wang, Zhou Su, Ning Zhang, Jianfei Chen, Xin Sun, Zhiyuan Ye, and Zhenyu Zhou.
\newblock Spds: A secure and auditable private data sharing scheme for smart grid based on blockchain.
\newblock {\em IEEE Transactions on Industrial Informatics}, 17(11):7688--7699, 2020.

\bibitem{ERC-721}
Ethereum Community.
\newblock Erc-721 non-fungible token standard, 2017.
\newblock URL: \url{https://ethereum.org/en/developers/docs/standards/tokens/erc-721/}.

\bibitem{IPFS}
IPFS Community.
\newblock Ipfs: An open system to manage data without a central server, 2024.
\newblock Accessed: 2024-10-15.
\newblock URL: \url{https://ipfs.tech/}.

\bibitem{ERC-1155}
Ethereum Community.
\newblock Erc-1155 multi-token standard, 2024.
\newblock URL: \url{https://ethereum.org/en/developers/docs/standards/tokens/erc-1155/}.

\end{thebibliography}

\end{document}